\documentclass[12pt, a4paper]{article}
\pdfoutput=1
\usepackage{color}
\usepackage{graphicx}
\usepackage{caption}
\usepackage{amsmath}
\usepackage{subcaption}
\usepackage{mathtools}
\usepackage{url}
\usepackage[font=scriptsize,labelfont=bf,skip=6pt]{caption}
\usepackage[T1]{fontenc}
\usepackage{authblk}

\begin{document}
\title{Development, Characterization and Qualification of first GEM foils produced in India}
\author{Aashaq Shah}
\author{Asar Ahmed}
\author{Mohit Gola}
\author{Ram krishna Sharma\thanks{Corresponding Author: rasharma@cern.ch}}
\author{Shivali Malhotra}
\author{Ashok Kumar}
\author{Md. Naimuddin}
\affil{Department of Physics $\&$ Astrophysics, University of Delhi, Delhi, India}
\author{Pradeep Menon}
\author{K. Srinivasan}
\affil{Micropack Pvt. Ltd., Bengaluru, India}
\maketitle
\begin{abstract}

The increasing demand for Gas Electron Multiplier (GEM) foils has been driven by their application in many current and proposed high-energy physics experiments. Micropack, a Bengaluru-based company, has established and commercialized GEM foils for the first time in India. Micropack used the double-mask etching technique to successfully produce 10 cm $\times$ 10 cm GEM foil. In this paper, we report on the development as well as the geometrical and electrical properties of these foils, including the size uniformity of the holes and leakage current measurements. Our characterization studies show that the foils are of good quality and satisfy all the necessary quality control criteria. 
\end{abstract}
\section{Introduction}
The concept of a GEM was introduced by F. Sauli in 1997 \cite{one}. GEM foils consist of a 50 $\mu$m thin polyimide (PI) foil coated with a thin layer of copper on both sides. Bi-conical holes with 50 $\mu$m inner and 70 $\mu$m outer diameters are chemically etched in the foil at a pitch of about 140$\mu$m by using either a double mask \cite{one} or single mask \cite{one_01} technique. GEM foils have attracted significant interest from the nuclear and particle physics communities, as they are excellent candidates to be used in tracking detectors. This detector technology has been used successfully as a tracking detector in many experiments, such as STAR \cite{two}, TOTEM \cite{two_01}, LHCb \cite{two_02}, COMPASS \cite{three} and ALICE \cite{four}, and is expected to be used in many future experiments and their upgrades \cite{five}. Presently, CERN is the main distributor of small as well as large area GEM foils. It is quite difficult for such a production site to meet the growing demands. To meet the future requisites, Micropack Pvt. Ltd. \cite{six} India has acquired a license from CERN to manufacture and commercialize GEM foils \cite{six_01}. Currently, Micropack has successfully produced double-mask GEM foils of 10 cm $\times$ 10 cm size. In this paper, we will describe the technique used for the foil production, details of the quality control tests and various characterization studies performed to validate the foils in order for them to be used for various applications.
\section{Foil Production}
Several Indian Institutions, including the University of Delhi, are part of the muon detector upgrade project of the CMS experiment \cite{six_02} at the Large Hadron Collider (LHC). Indian groups are planning to contribute approximately 20 $\%$ of the total GEM detectors required for the CMS GE1$\slash$1 upgrade \cite{six_03} and future GE2$\slash$1 upgrade \cite{six_04} . As a result, an intensive R$\&$D program on GEM detectors has been initiated at these institutions. Micropack Pvt. Ltd. in Collaboration with Indian Institutions have embarked upon the development of GEM foils in India.

In the later part of 2013, Micropack signed a Transfer of Technology (TOT) agreement with CERN for the development of GEM foils in India. After continuous efforts, refining of processes and repeated trials, Micropack has been successful in realizing 10 cm $\times$ 10 cm GEM foils, meeting the standard dimensional requirements. The foil production at Micropack started with single-mask process. However, after several attempts it was realized that copper removal through the reverse plating method was challenging. By contrast, the double-mask process succeeded and gave fast results. The double-mask GEM foils were produced by Micropack in a similar fashion as produced at the CERN PCB workshop \cite{six_05}, using photo-lithographic techniques in which hole patterns are transferred to the copper-clad polyimide substrate using microscopic masks placed on the top and bottom of the substrate. A 15 $\mu$m thick photo-resistive layer is applied on both sides of the substrate and the mask is placed on top of the base material and engraved on the photo-resist by UV-light exposure. The foil used was a 50 $\mu$m PI (Apical Type NP) film with 5 $\mu$m copper foil on either side. Several solvents and acid baths are used to etch copper layer  to form the copper holes. The polyimide is then dissolved by chemical etching using the copper layer as a mask. 
\begin{figure}[!ht]
    \centering
    \begin{subfigure}[b]{0.46\textwidth}
        \includegraphics[width=6cm, height=4cm]{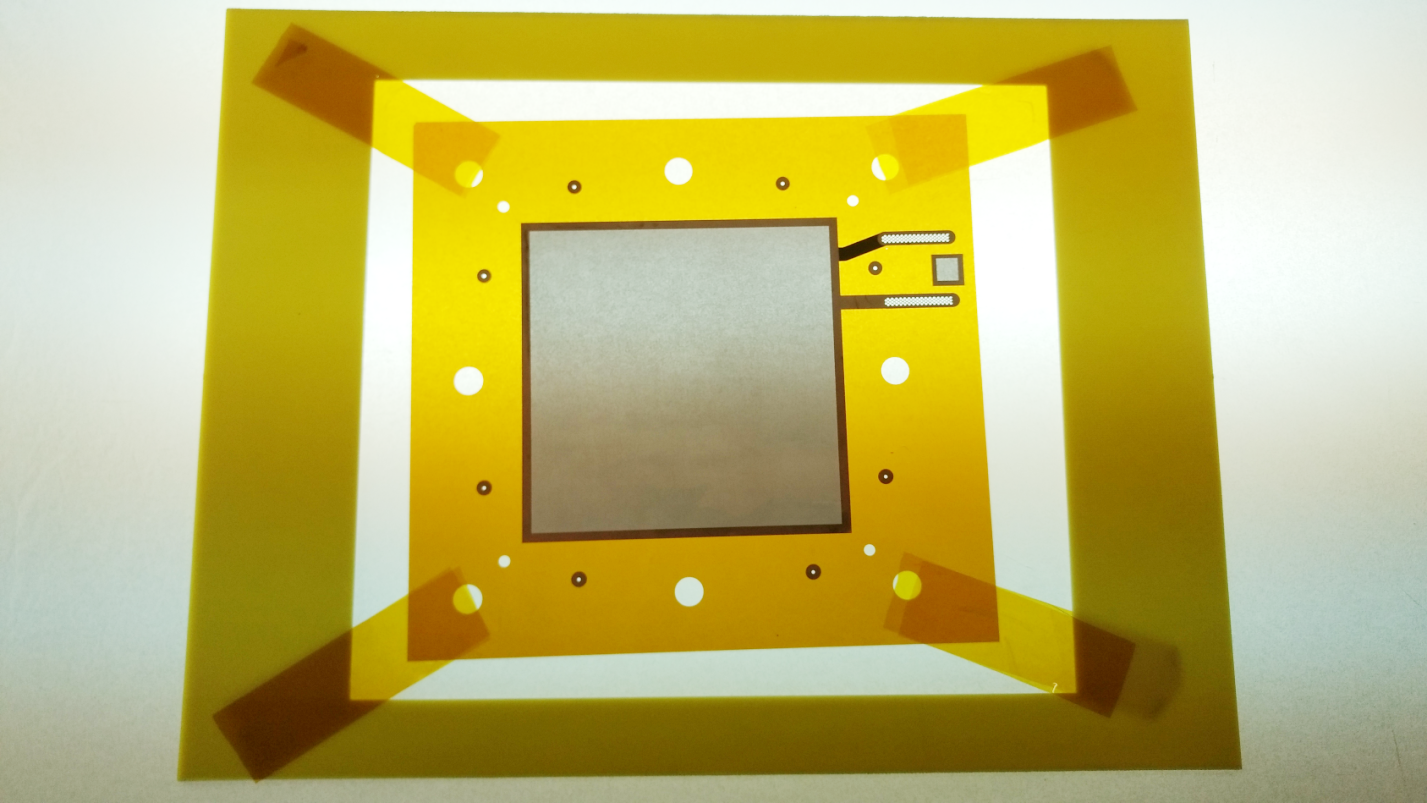}\qquad
        \caption{ }
    \end{subfigure}
    \begin{subfigure}[b]{0.46\textwidth}
        \includegraphics[width=6cm, height=4cm]{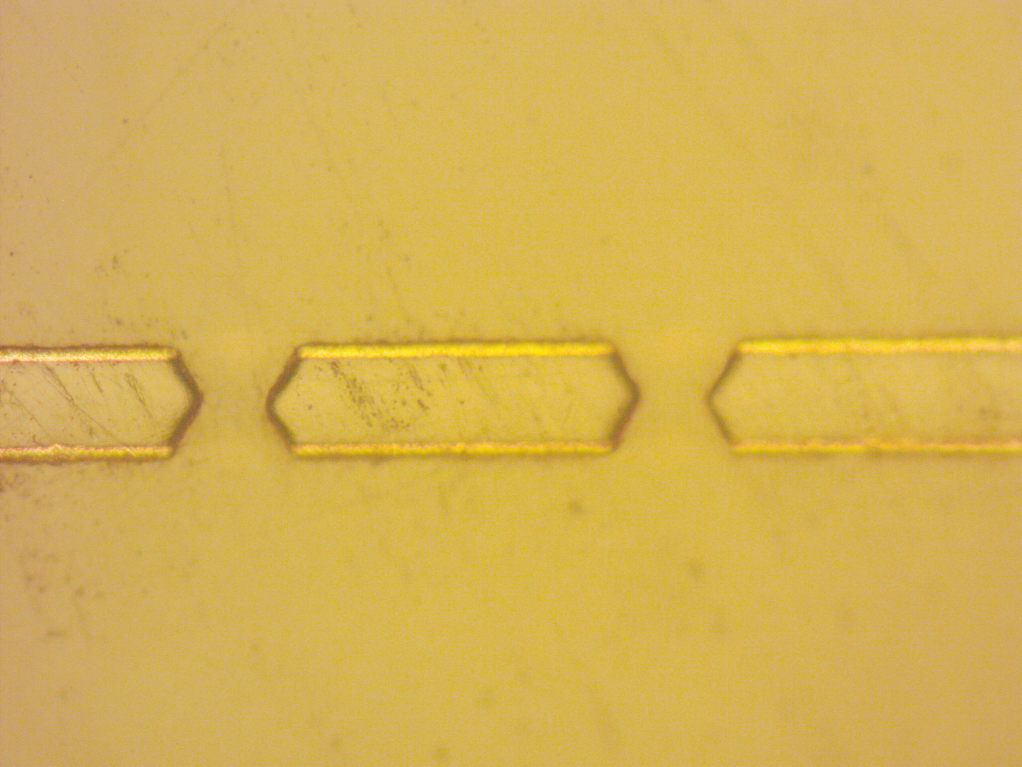}
        \caption{ }
    \end{subfigure}
   \caption{(a) 10 cm $\times$ 10 cm GEM foil encapsulated in a frame and (b) Cross-sectional view of the foil showing the double cone structure of the engraved holes. } \label{fig:Foil_and_Cone}
\end{figure}

Figure \ref{fig:Foil_and_Cone} (a) shows the newly produced 10 cm $\times$ 10 cm GEM foil. Figure \ref{fig:Foil_and_Cone} (b) shows the cross-sectional view of the foil showing the double cone structure of the engraved holes. The realization of the foils has been achieved primarily through accurate lithographic and controlled chemical processes with a double cone hole structure to enhance the end gain.

In order to qualify these GEM foils as commercially and scientifically reliable, a number of quality control tests needed to be performed. Therefore, we have characterized the foils by studying their optical and electrical properties to render them usable for further applications.
\subsection{Optical Assessment}

\begin{figure}[!ht]
    \centering
        \includegraphics[width=12cm, height=8cm]{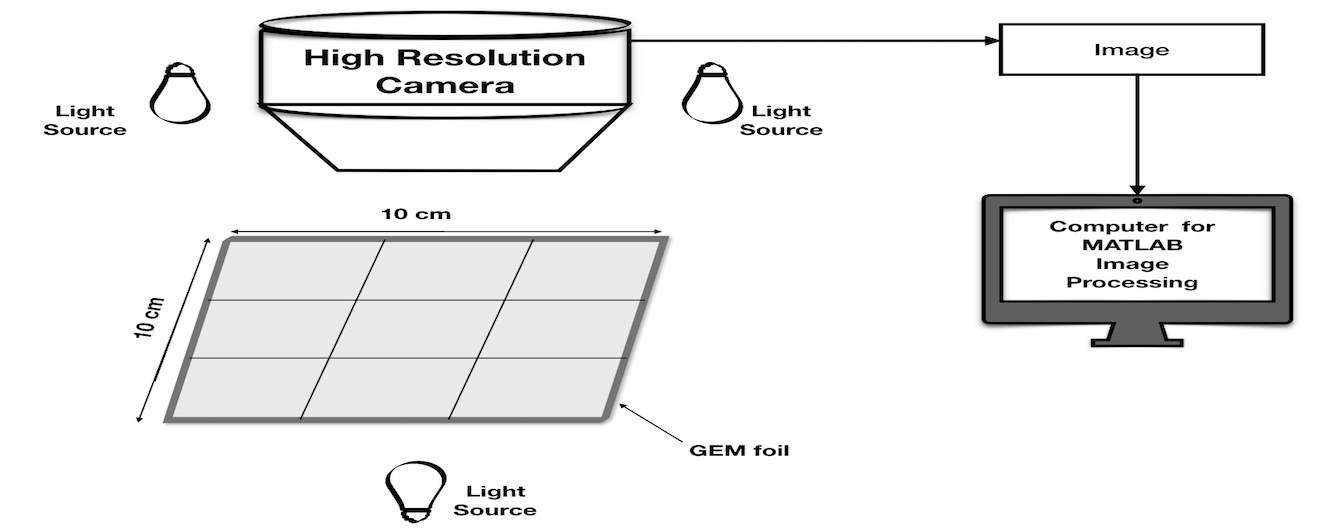}
   \caption{Sketch of the setup used for the optical measurements.}   \label{fig:Optical_Sketch}
\end{figure}

The GEM foil performance depends heavily upon the hole geometry and their pattern. A GEM foil with a 140 $\mu$m pitch using a hexagonal hole pattern contains approximately 600,000 holes. Any irregularity or defect in the hole pattern and its geometry can profoundly affect their performance. It becomes therefore very important to study the hole geometry structure of the foil and to locate every glitch and piece of debris which could lead to foil failure. Though the qualitative estimate of hole density and diameters can  manually be studied using optical microscope but such a technique can become labor intensive especially when there are large number of holes to be analyzed. To overcome this problem, various techniques have been developed \cite{eight, nine} to study the optical properties, where geometrical properties of the foils have been measured using an automated 2D CCD scanner. However, in our study we have used a slightly different approach to explore the geometrical properties of the GEM foils. Each of the foils were scanned using Micro lensing technique with an AF-S Micro Nikon 40 mm 1:2.8G lens where multiple images of micrometer resolution per pixel were captured. A Soft Box (1 m $\times$ 1 m) light source has been used to provide uniform illumination to the GEM foils. A sketch of the optical measurement setup is shown in the Figure~\ref{fig:Optical_Sketch}.

\begin{figure}[!ht]
    \centering
    \begin{subfigure}[b]{0.29\textwidth}
        \includegraphics[width=4cm, height=3cm]{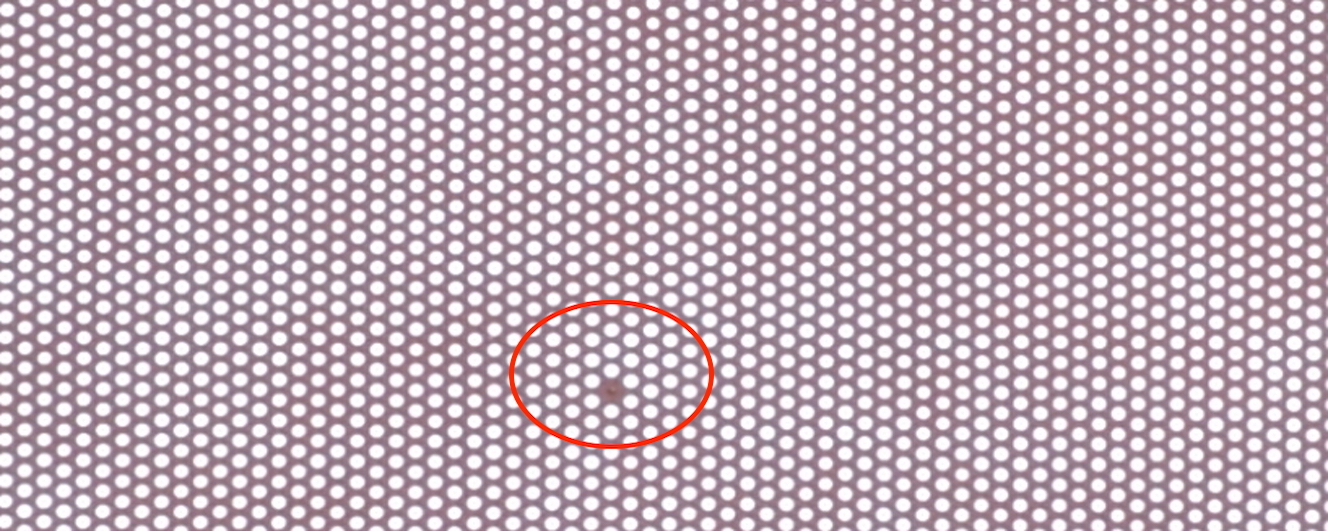}
        \caption{ }
        \label{fig:O_4a}
    \end{subfigure}
    \begin{subfigure}[b]{0.29\textwidth}
        \includegraphics[width=4cm, height=3cm]{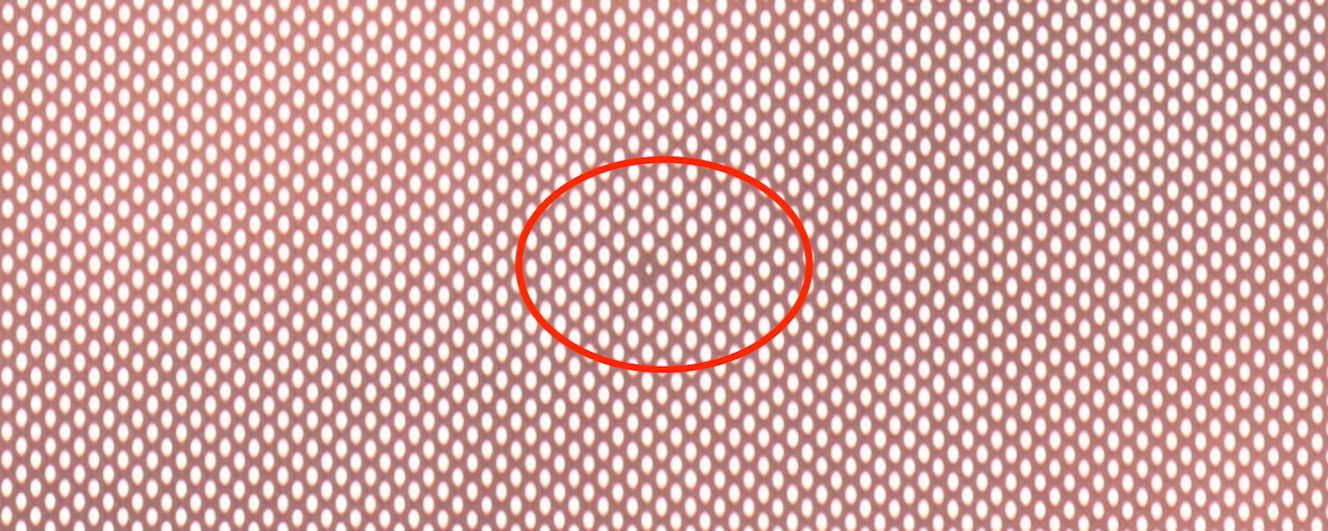}
        \caption{ }
        \label{fig:O_4b}
    \end{subfigure}
    \centering
    \begin{subfigure}[b]{0.29\textwidth}
        \includegraphics[width=4cm, height=3cm]{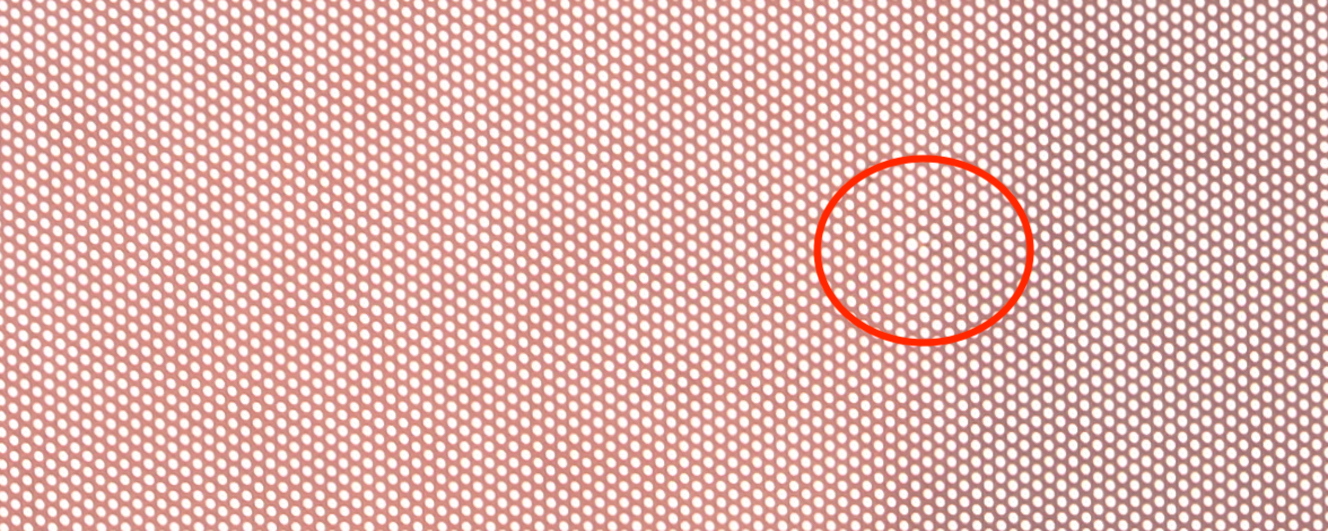}
        \caption{ }
        \label{fig:O_4c}
    \end{subfigure}
    \centering
    \begin{subfigure}[b]{0.29\textwidth}
        \includegraphics[width=4cm, height=3cm]{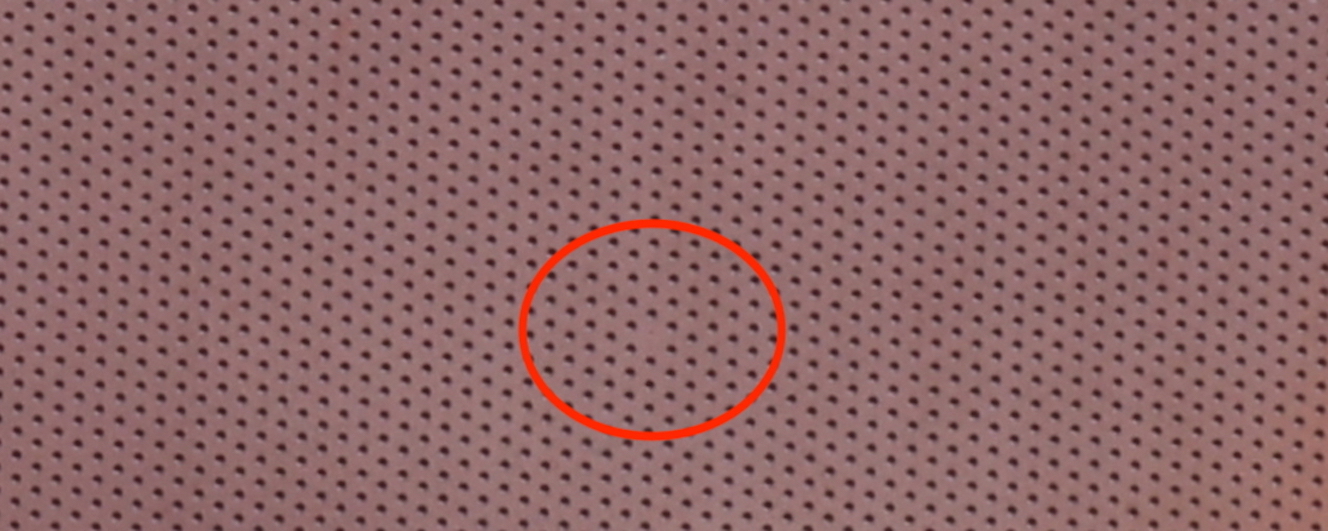}
        \caption{ }
        \label{fig:O_5a}
    \end{subfigure}
    \centering
    \begin{subfigure}[b]{0.29\textwidth}
        \includegraphics[width=4cm, height=3cm]{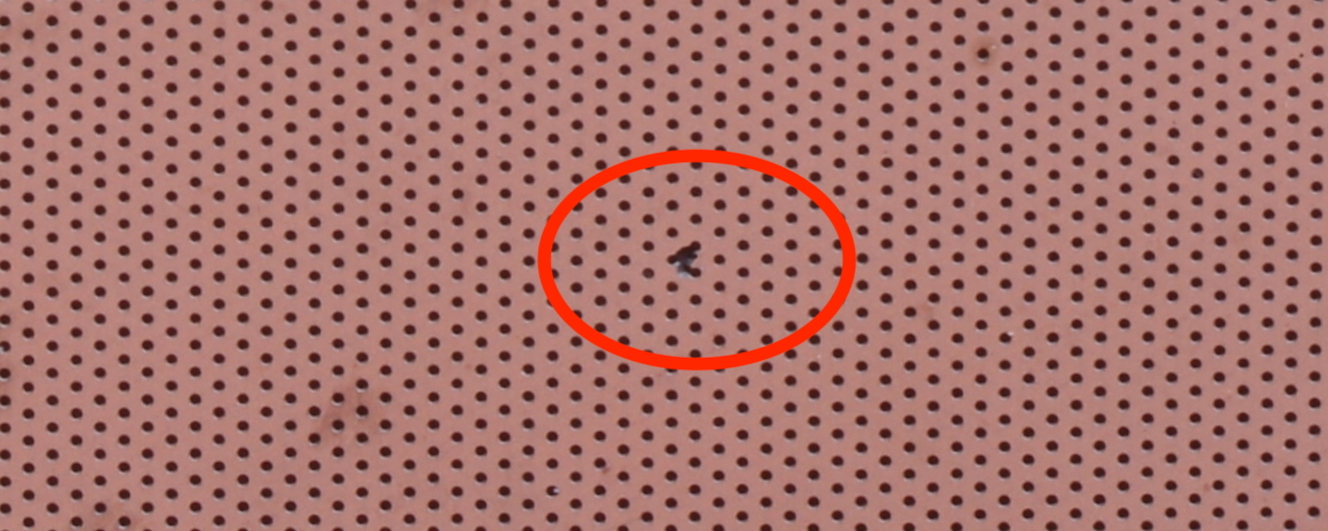}
        \caption{ }
        \label{fig:O_5b}
    \end{subfigure}
    \centering
    \begin{subfigure}[b]{0.29\textwidth}
        \includegraphics[width=4cm, height=3cm]{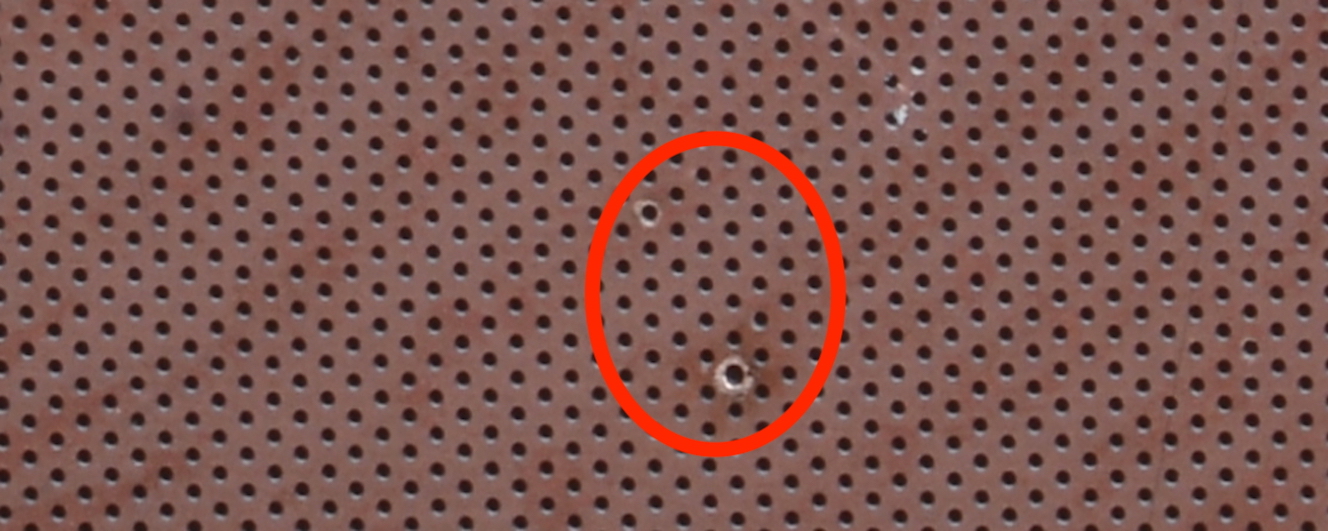}
        \caption{ }
        \label{fig:O_5c}
    \end{subfigure}
   \caption{Observed imperfections in the foils: (a) Un-etched area, (b) under-size hole, (c) over-size hole (d) missing hole, (e) excess etching and (f) burnt area.} \label{fig:Optical_01}
\end{figure}

The quantities that have been optically measured are the inner and outer hole diameters. The various kinds of possible imperfections that have been observed are un-etched areas, under-size hole, oversize holes, without hole areas, excess etching and burnt holes. All these imperfections are shown in Figure \ref{fig:Optical_01}. Also, the scan with the front light ON and the back light OFF has been performed as to make the scan sensitive to the outer holes. For the inner holes of the foil, the scan has been performed with the front light OFF and back light ON.
\begin{figure}[!ht]
    \begin{subfigure}[b]{0.40\textwidth}
        \includegraphics[width=7cm, height=5.5cm]{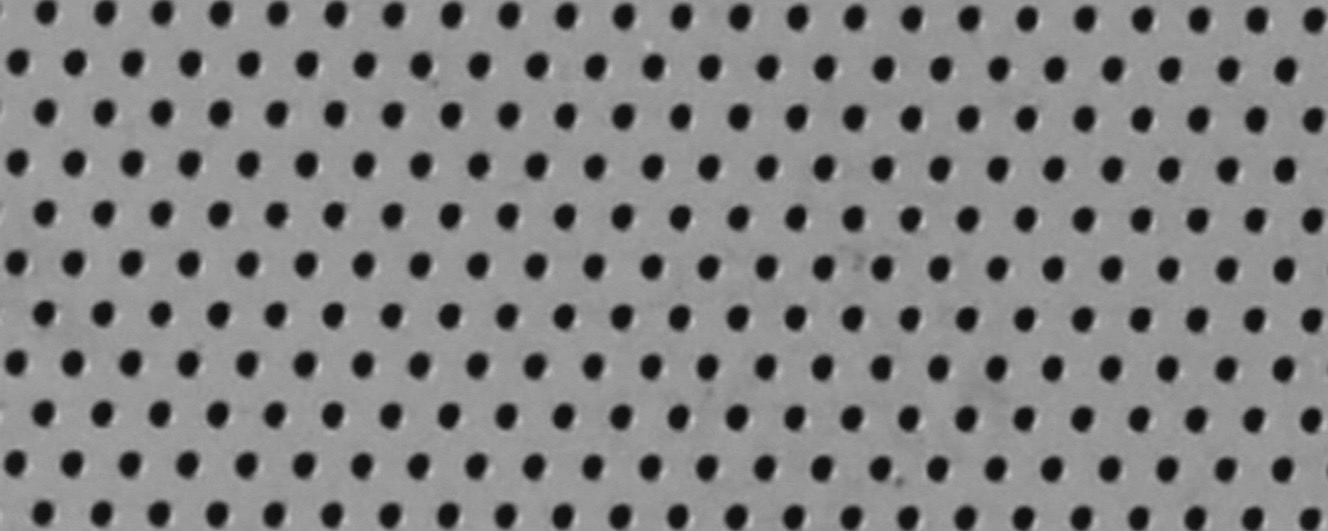}
        \caption{ }
        \label{fig:O_2}
    \end{subfigure}
    \begin{subfigure}[b]{0.40\textwidth}
        \includegraphics[width=6.5cm, height=5.5cm]{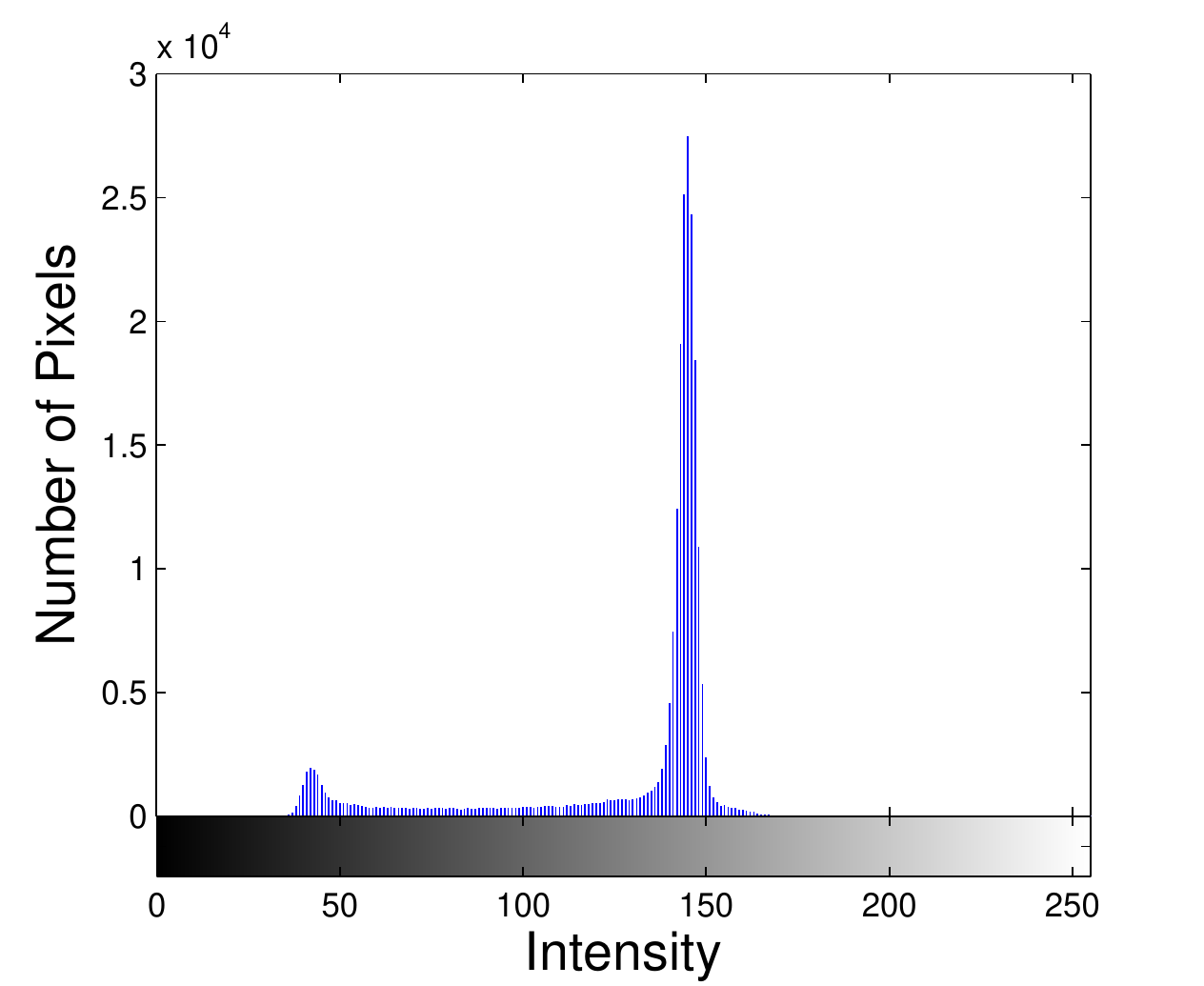}
        \caption{ }
        \label{fig:O_3}
    \end{subfigure}
   \caption{(a) Image formed in gray-scale (b) Histogram of gray-scale image for the calculation of gray threshold.} \label{fig:Optical_03}
\end{figure}
\begin{figure}[!ht]
    \centering
    \begin{subfigure}[b]{0.52\textwidth}
        \includegraphics[width=7cm, height=5cm]{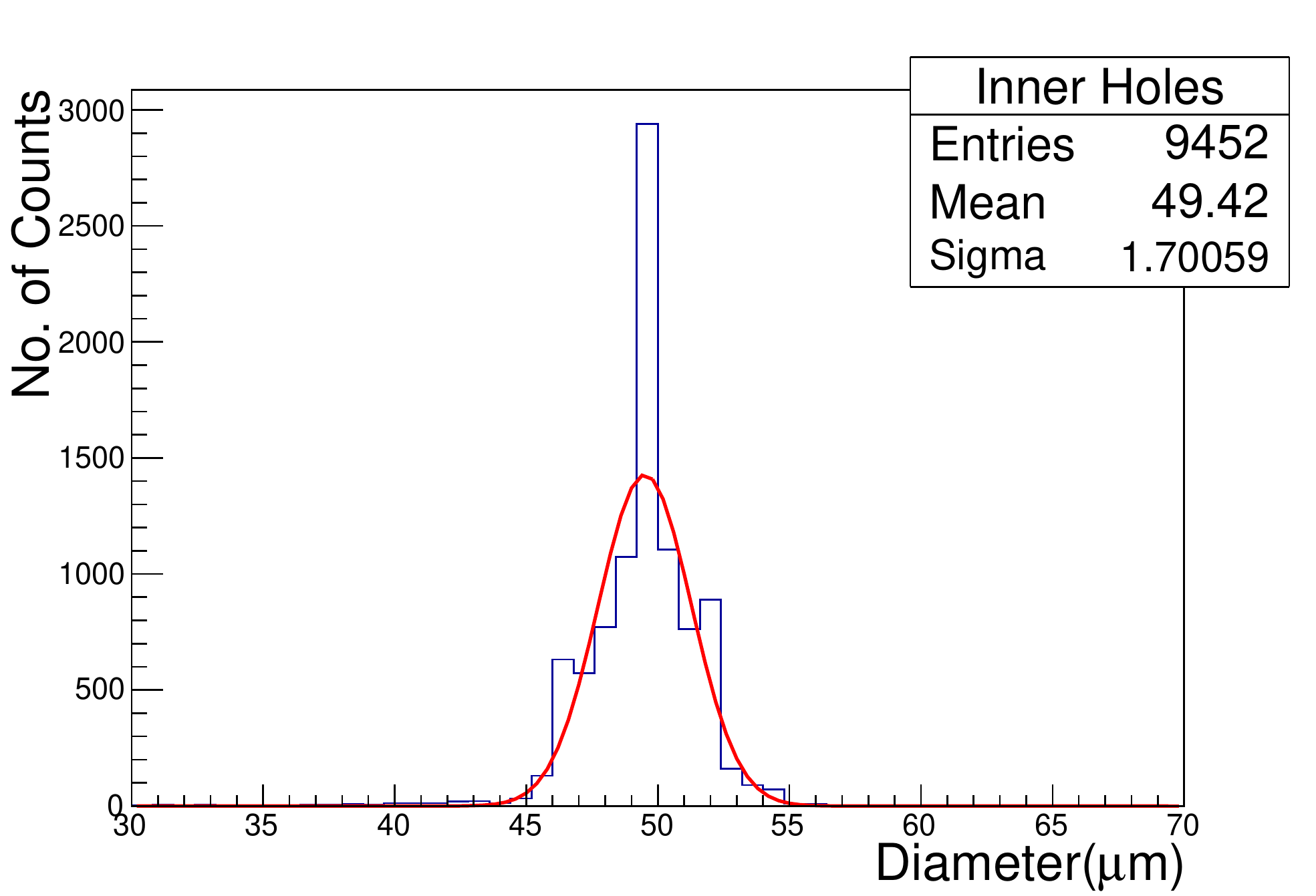}
        \caption{ }
        \label{fig:DiameterDistributionSector_inner}
    \end{subfigure}
    \begin{subfigure}[b]{0.45\textwidth}
        \includegraphics[width=7cm, height=5cm]{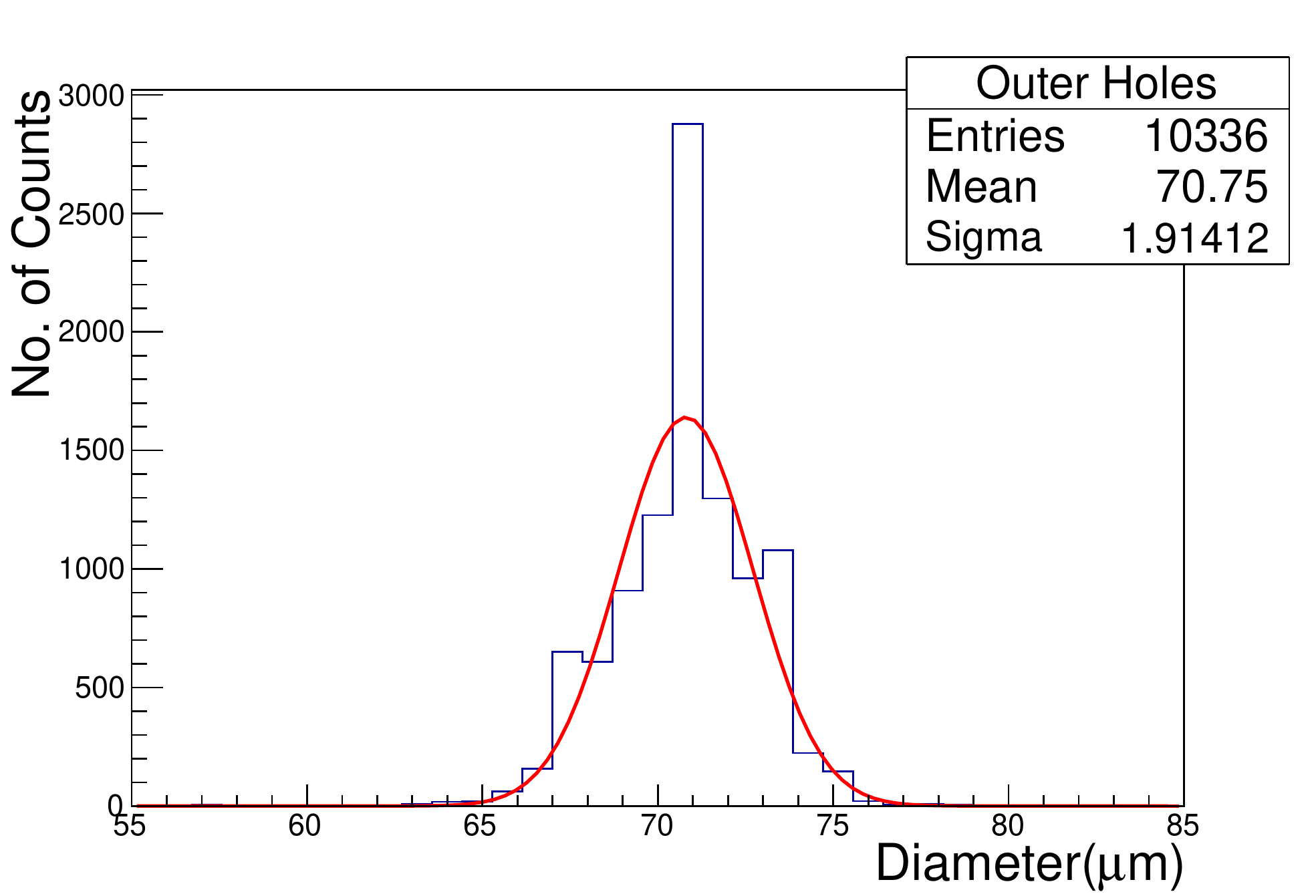}
        \caption{ }
        \label{fig:DiameterDistributionSector_outer}
    \end{subfigure}
   \caption{Hole size distribution of (a) inner, and  (b) outer holes for one sector.} \label{fig:DiameterDistributionSector}
\end{figure}
\begin{figure}[!ht]
    \centering
    \begin{subfigure}[b]{0.52\textwidth}
        \includegraphics[width=7cm, height=5cm]{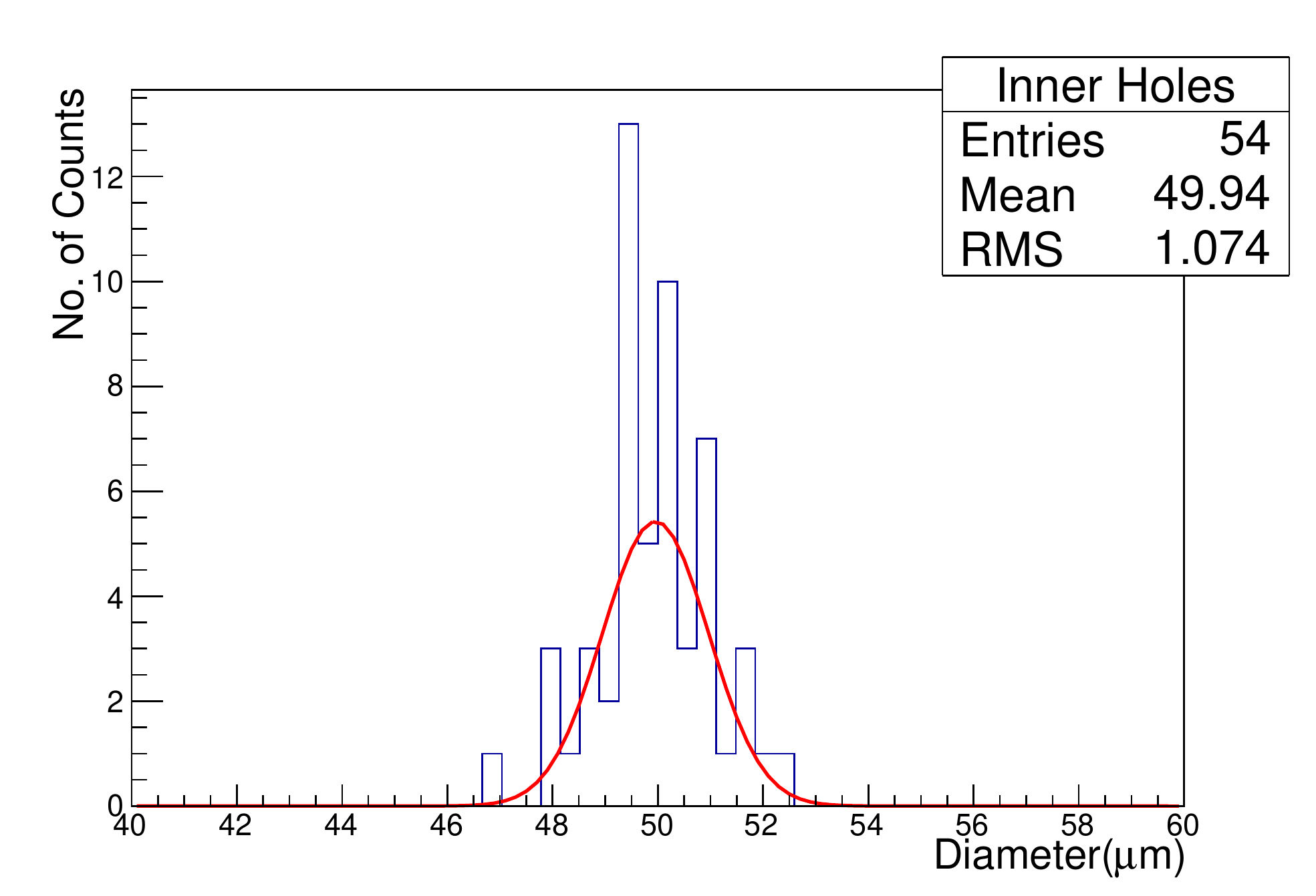}
        \caption{ }
        \label{fig:O_6}
    \end{subfigure}
    \begin{subfigure}[b]{0.47\textwidth}
        \includegraphics[width=7cm, height=5cm]{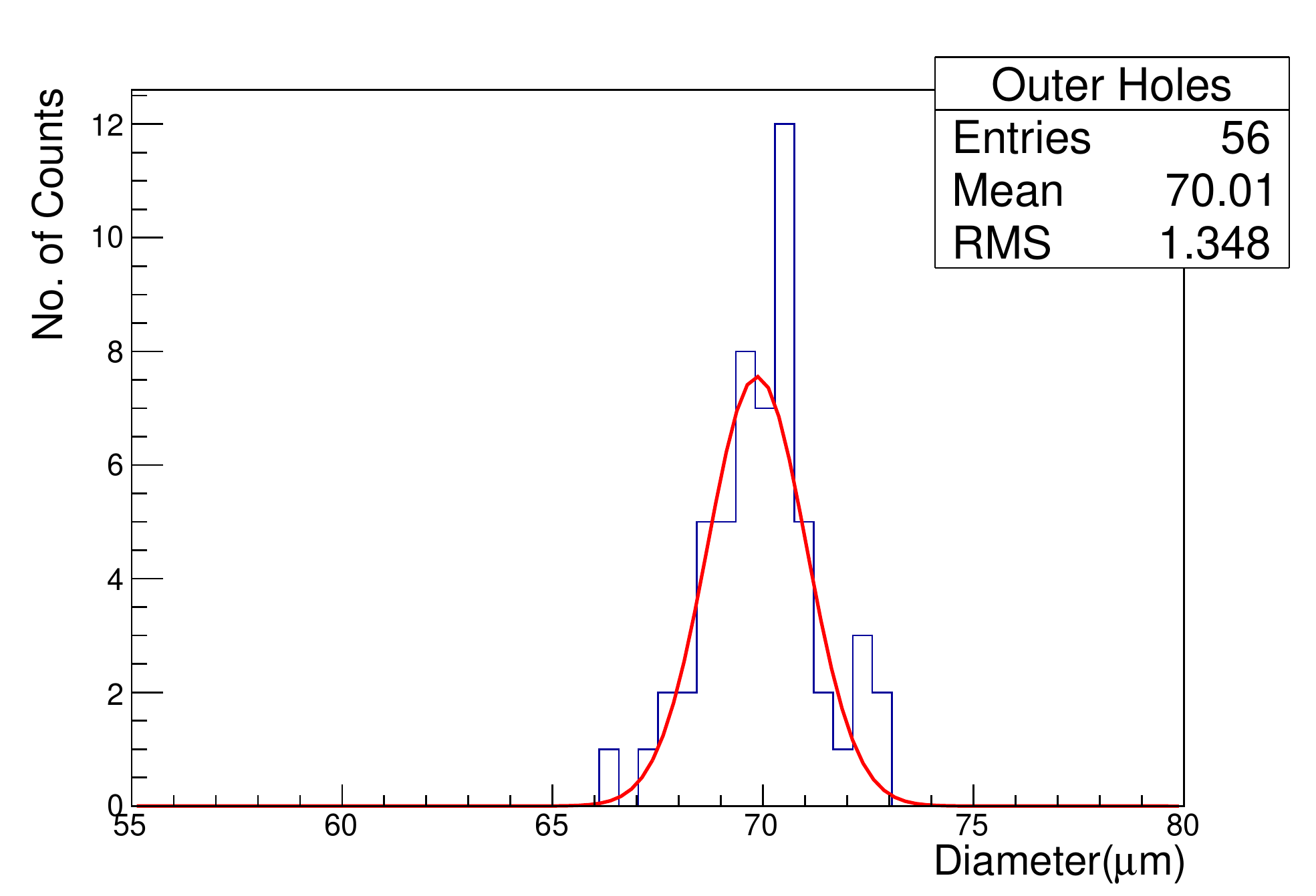}
        \caption{ }
        \label{fig:O_7}
    \end{subfigure}
   \caption{Mean diameter of (a) Inner holes of all the sectors (b) Outer holes of all the sectors. The hole distributions were fitted with Gaussian functions to extract values for mean and standard deviation as shown in Figure \ref{fig:Optical_05}. } \label{fig:Mean_diameter}
\end{figure}

\begin{figure}[!ht]
    \centering
    \begin{subfigure}[b]{0.46\textwidth}
        \includegraphics[width=6.5cm, height=5cm]{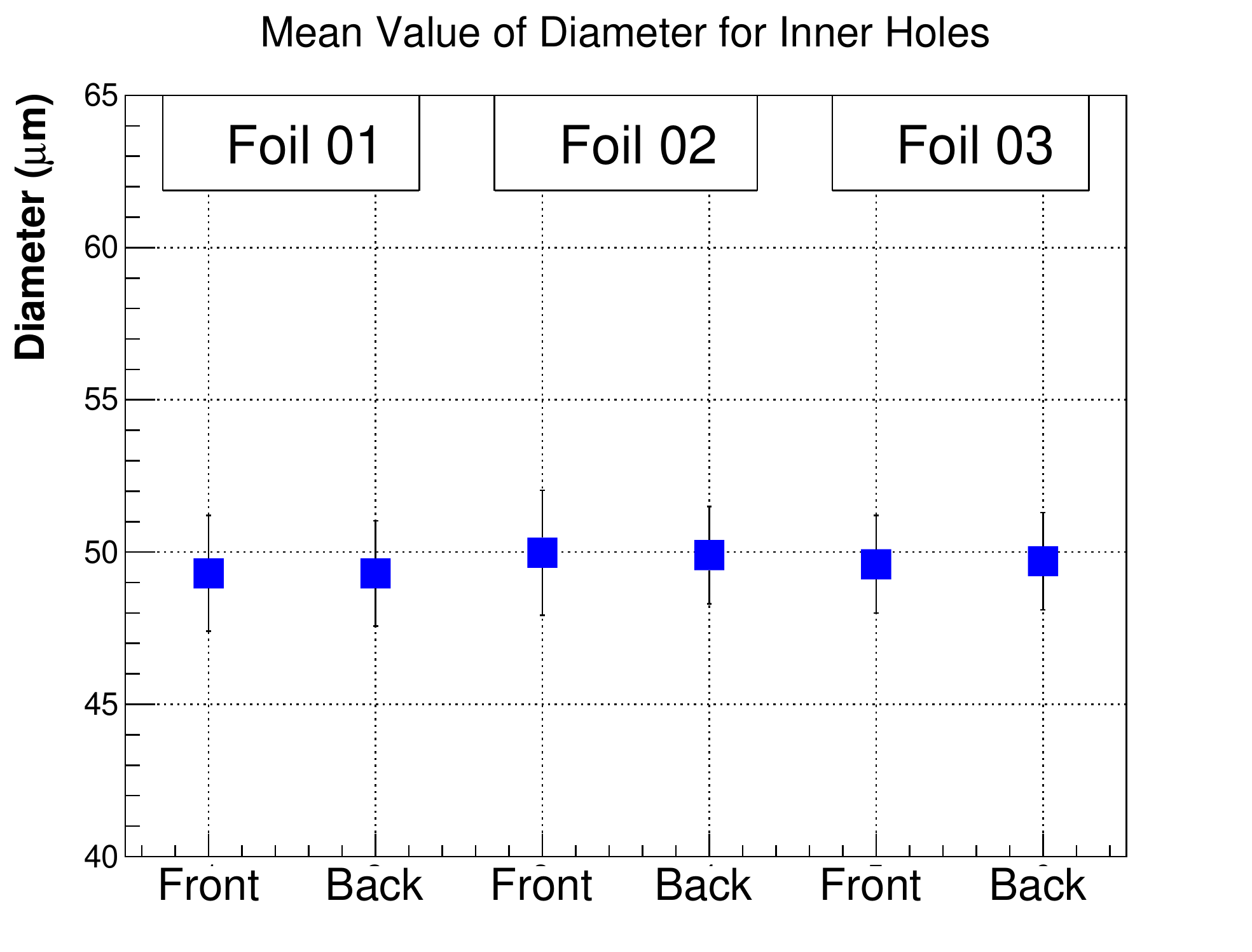}\qquad
        \caption{ }
        \label{fig:O_8a}
    \end{subfigure}
    \begin{subfigure}[b]{0.46\textwidth}
        \includegraphics[width=6.5cm, height=5cm]{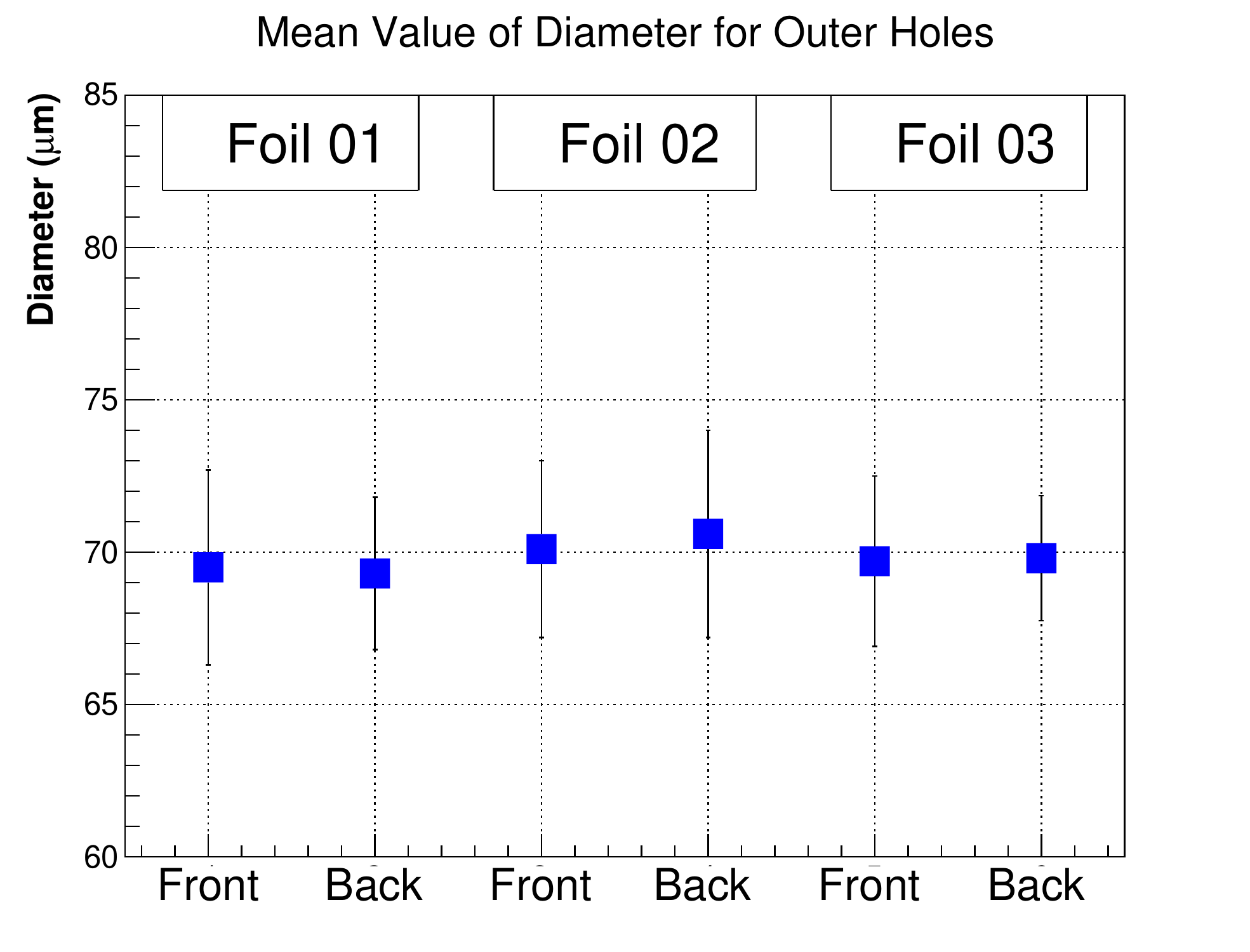}
        \caption{ }
        \label{fig:O_8b}
    \end{subfigure}
   \caption{Mean diameter for (a) Inner and (b) outer holes for each side of GEM foils. The error bars represent the 1 standard deviation error obtained from statistical combination of the standard deviations of hole diameter distributions of each sub-sector.} \label{fig:Optical_05}
\end{figure}

\begin{figure}[!ht]
    \centering
    \begin{subfigure}[b]{0.45\textwidth}
        \includegraphics[width=6.5cm, height=5cm]{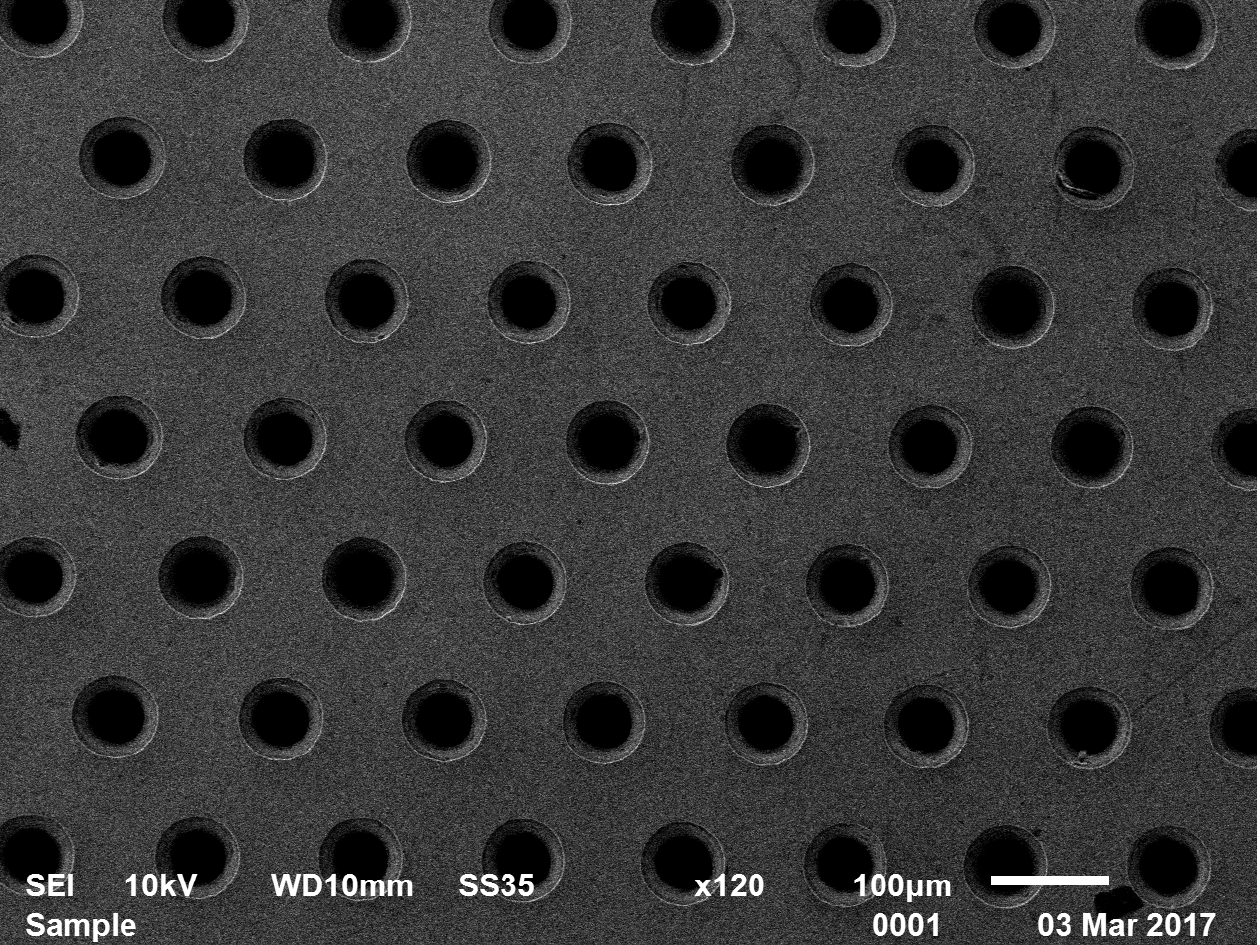}\qquad
        \caption{ }
        \label{fig:SEM_01}
    \end{subfigure}
    ~ 
    \begin{subfigure}[b]{0.45\textwidth}
        \includegraphics[width=6.5cm, height=5cm]{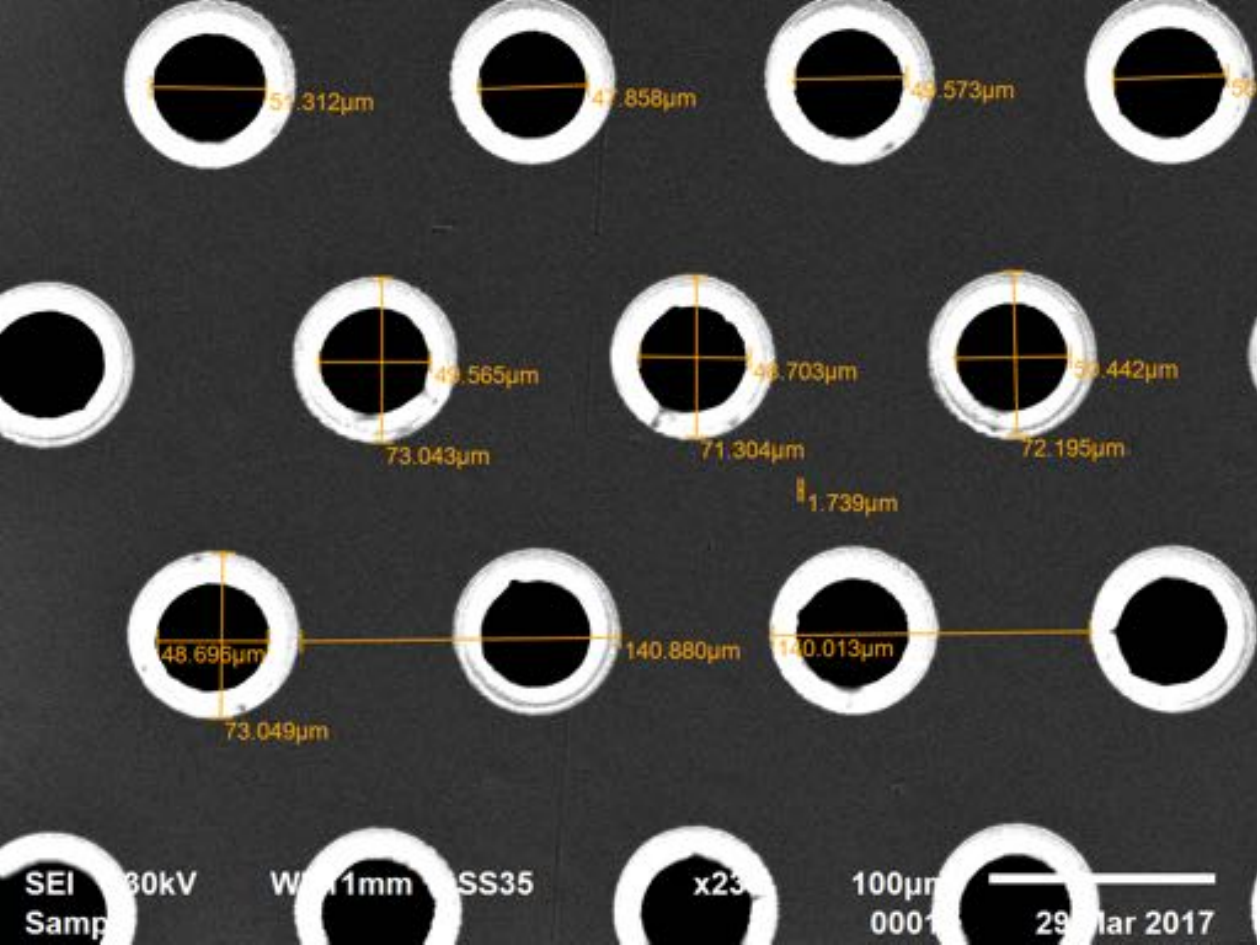}
        \caption{ }
        \label{fig:SEM_03}
    \end{subfigure}
    ~ 
    \caption{(a) SEM image at $\mu$m level resolution showing the overall uniformity of the foil sample and (b) hole diameters and the pitch under SEM  at $\mu$m level resolution.}\label{fig:SEM}
\end{figure}

\begin{figure}[!ht]
    \centering
    \begin{subfigure}[b]{0.49\textwidth}
        \includegraphics[width=7.6cm, height=5.5cm]{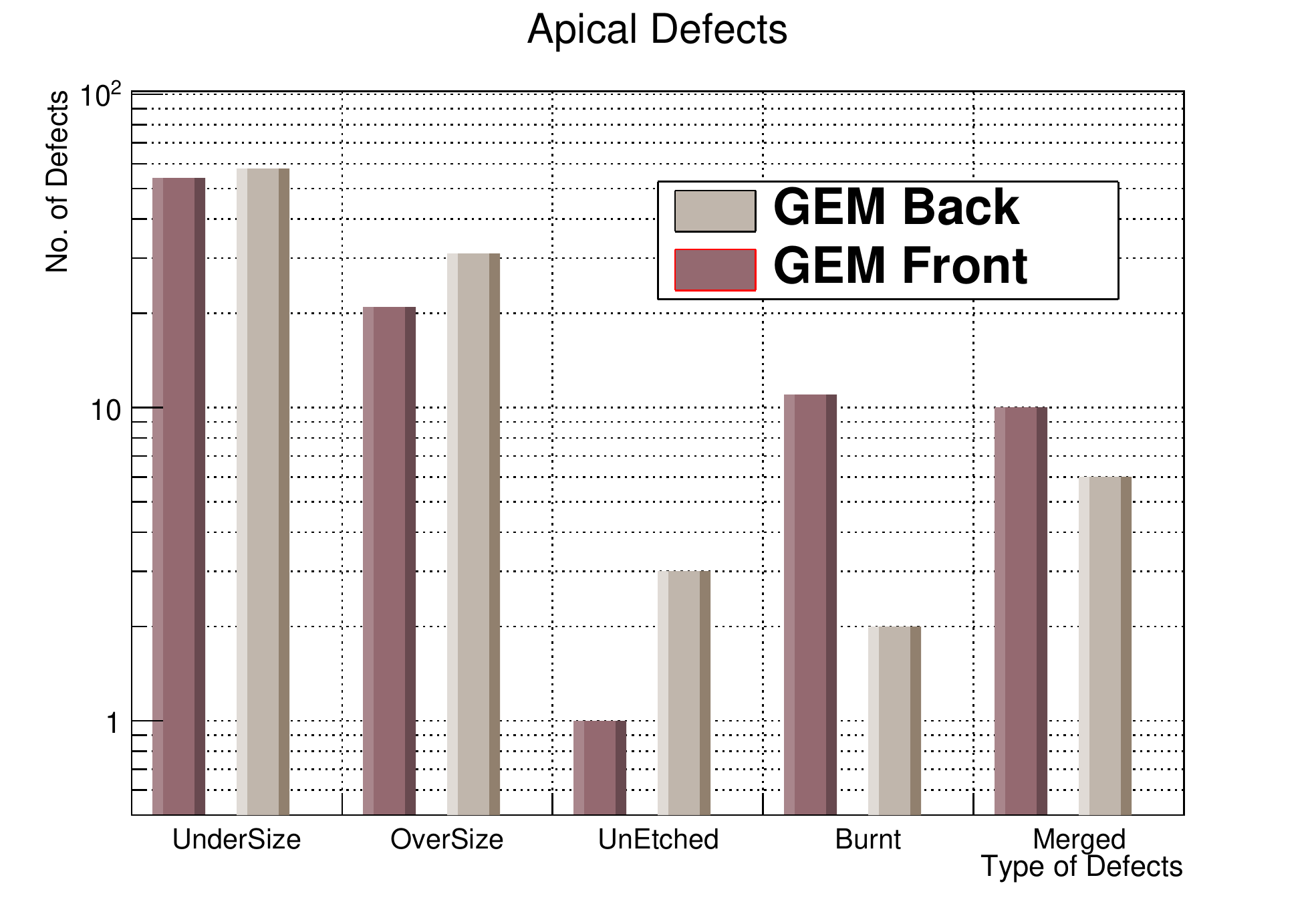}\qquad
        \caption{ }
        \label{fig:O_9a}
    \end{subfigure}
    \begin{subfigure}[b]{0.49\textwidth}
        \includegraphics[width=7.6cm, height=5.5cm]{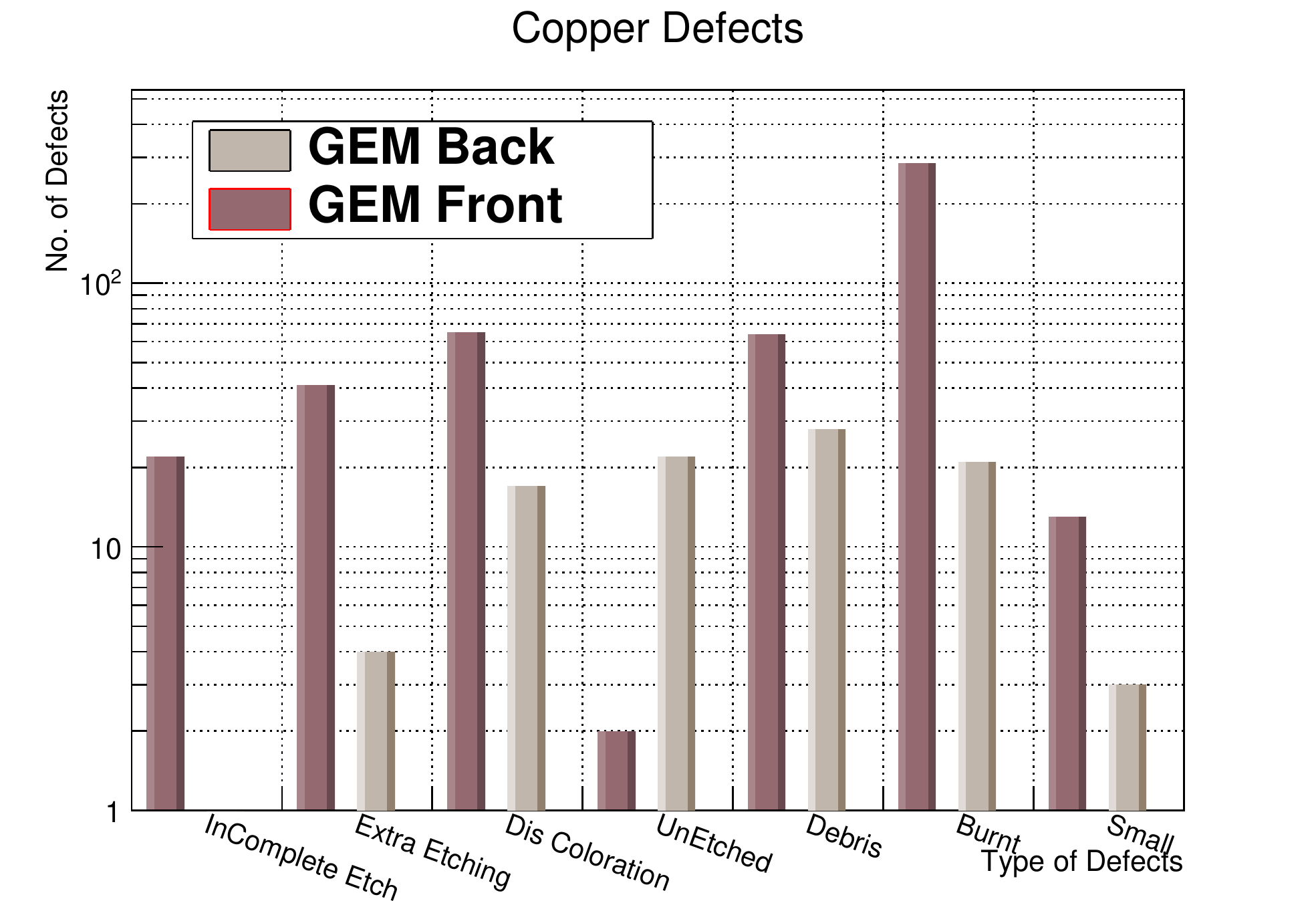}
        \caption{ }
        \label{fig:O_9b}
    \end{subfigure}
   \caption{Number of defects seen in (a) Insulator (Apical Type NP) and (b) Copper, for one of the 10 cm $\times$ 10 cm foil.} \label{fig:Optical_04}
\end{figure}


To assess the entire area of the foil, each of the foil were divided into several sub-sectors. While capturing inner holes, the foil was divided into 54 (9$\times$6) sub-sectors, whereas to capture outer holes the foil was divided into 56 (8$\times$7) sub-sectors. Images were captured in such a way that each image corresponds to a sub-sector. Each captured image has been then processed using an Image Processing Toolkit within MATLAB \cite{ten_01}, which contains built-in functions specifically designed to convert the pixel information obtained from images into numerical measurements. Therefore, the toolkit has been used to convert the primary image acquired by the camera into gray-scale image as shown in Figure \ref{fig:Optical_03} (a). In order to obtain the binary threshold to separate the holes from background, the gray scale image was converted into intensity histogram as shown in Figure \ref{fig:Optical_03} (b). The left peak in the figure represents the light reflected back from the edges of holes and screen behind the foil while the right peak represents the light from the copper surface. Each of the holes were labeled for each sub-sector. The hole diameter in pixels were calculated for each side of the GEM foils. 

The data generated from MATLAB has been processed further using ROOT \cite{ten_02} to estimate the mean diameter in pixels of inner and outer holes for each sub-sector and finally for the entire foil. The diameter values in pixels were converted into micrometers using the image resolution of inner and outer holes as 5.6 $\mu$m/pixel and 7.2 $\mu$m/pixel, respectively. An example of the hole size distribution of inner and outer holes are shown for one of the sub-sectors in Figure~\ref{fig:DiameterDistributionSector}. From the fit to the distribution of holes diameter for each sub-sector, we obtained the mean diameter and standard deviation values for all the sub-sectors. As a result, we obtained 54 values of mean and sigma corresponding to 54 sub-sector for inner holes and 56 values of mean and sigma corresponding to 56 sub-sectors for the outer holes. We then statistically combine these individual means and sigmas of each sub-sector to estimate the mean diameter and standard deviation for inner and outer holes for the entire GEM foil. The distribution of mean diameters of all the sub-sectors for inner and outer holes of one GEM foil is shown in Figure \ref{fig:Mean_diameter}. The mean hole diameter for the entire foils estimated from Gaussian fit of this distribution gives a value of 49.9 $\mu$m and 70.01 $\mu$m for inner and outer holes respectively. The standard deviation obtained from each sub-sector has been statistically combined to extract the value over the entire foil and was found to be 1.6 $\mu$m and 2.02 $\mu$m for inner and outer holes respectively. The pitch obtained from the optical measurement is 140.0 $\pm$ 2.4 $\mu$m. The mean diameter of inner and outer holes for all the three foils are shown in Figure \ref{fig:Optical_05}. The error bars on the mean diameters shows the value of standard deviation. The findings are consistent with the double mask GEM foils produced else where and in use \cite{nine,ten}. Further, in the Figure \ref{fig:SEM} (a), scanning electron microscope (SEM) images of one of the GEM foil are shown, and Figure \ref{fig:SEM} (b) shows the average inner and outer hole diameters of 49.51 $\mu$m and 72.55 $\mu$m, respectively with an average pitch of 140.44 $\mu$m. This measurement of hole diameters from SEM measurement is in fair agreement with the values obtained from optical assessment.

The number of each type of defect in Apical Type NP or in Copper has been estimated and are shown in Figure \ref{fig:Optical_04}. There were a total of 785 number of defects including Copper and Apical Type NP out of approximately 600,000 holes in one of the 10 cm $\times$ 10 cm GEM foils which correspond to  0.13$\%$ of defects. Similar number of defects were also observed in the other two foils. Earlier optical studies \cite{ten_02a} on CERN foils have revealed similar defects. More recently, ALICE collaboration has also started an effort to optically characterize all the foils that they are planning to use for Time Projection Chamber (TPC) detector \cite{ten_02b, ten_02c}.

\subsection{Electrical Assessment}
The production quality of GEM foils can be quantified through optical and electrical tests. The optical test gives the information regarding the hole geometry and pitch related information whereas electrical test provides the parameters about the efficacy of the foils and hence is important in determining the quality of GEM foils. Electrical properties of the GEM foils were discerned by measuring its leakage current extended over a period of time after proper cleaning using adhesive roller. We divide electrical tests mainly in two types, quality control short or fast (QC fast) and quality control long (QC long) as per the CERN standards of quality control classification \cite{eleven}, which requires these two tests to be done in order to qualify these foils. The difference between QC fast and QC long lies in applying voltage for shorter or longer periods of time respectively, and monitoring the current. The other difference being that the QC fast gives the preliminary idea of leakage current or electrical connectivity of the foil but for more detailed study, QC long provides the behavior of the foil at high voltages in terms of information regarding the actual leakage current and the number of discharges, if any, for the reasonably longer duration of time. Here, both the tests have been performed; the electrical connectivity of the foils by QC fast method has been done with insulation tester MIT Megger 420 \cite{twelve}. Using this test, we established that the foils have good electrical connectivity.
\begin{figure}[!ht]
    \centering
        \includegraphics[width=12cm,height=8cm]{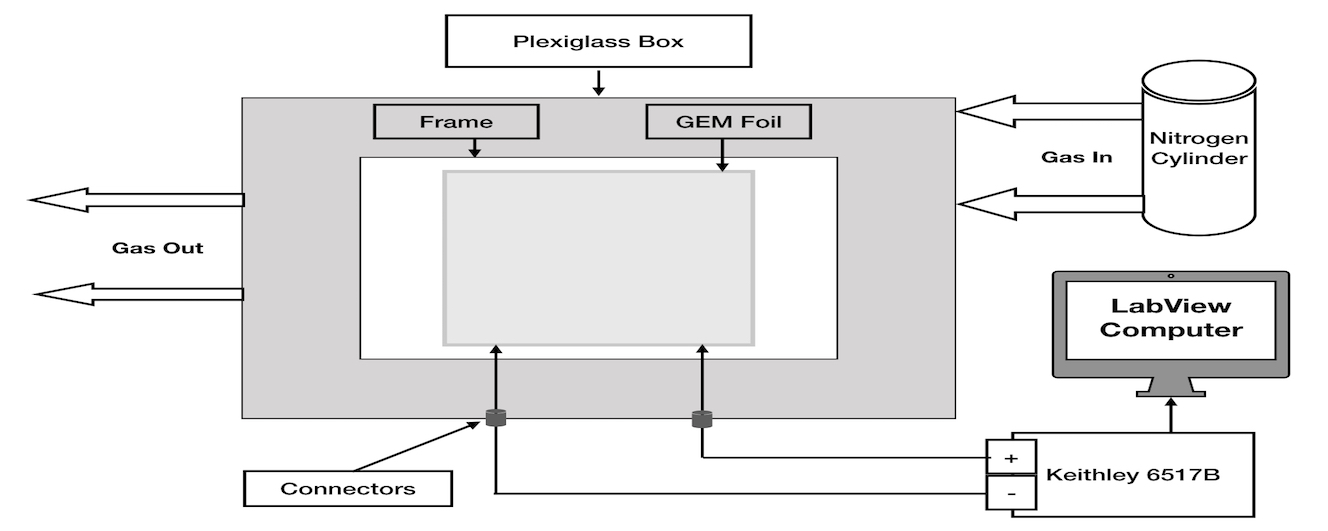}
   \caption{Sketch of the setup used for the measurement of leakage current.} \label{fig:Cleaning_Measurement}
\end{figure}
\begin{figure}[!ht]
    \centering
    \begin{subfigure}[b]{0.5\textwidth}
        \includegraphics[width=7.5cm, height=5.5cm]{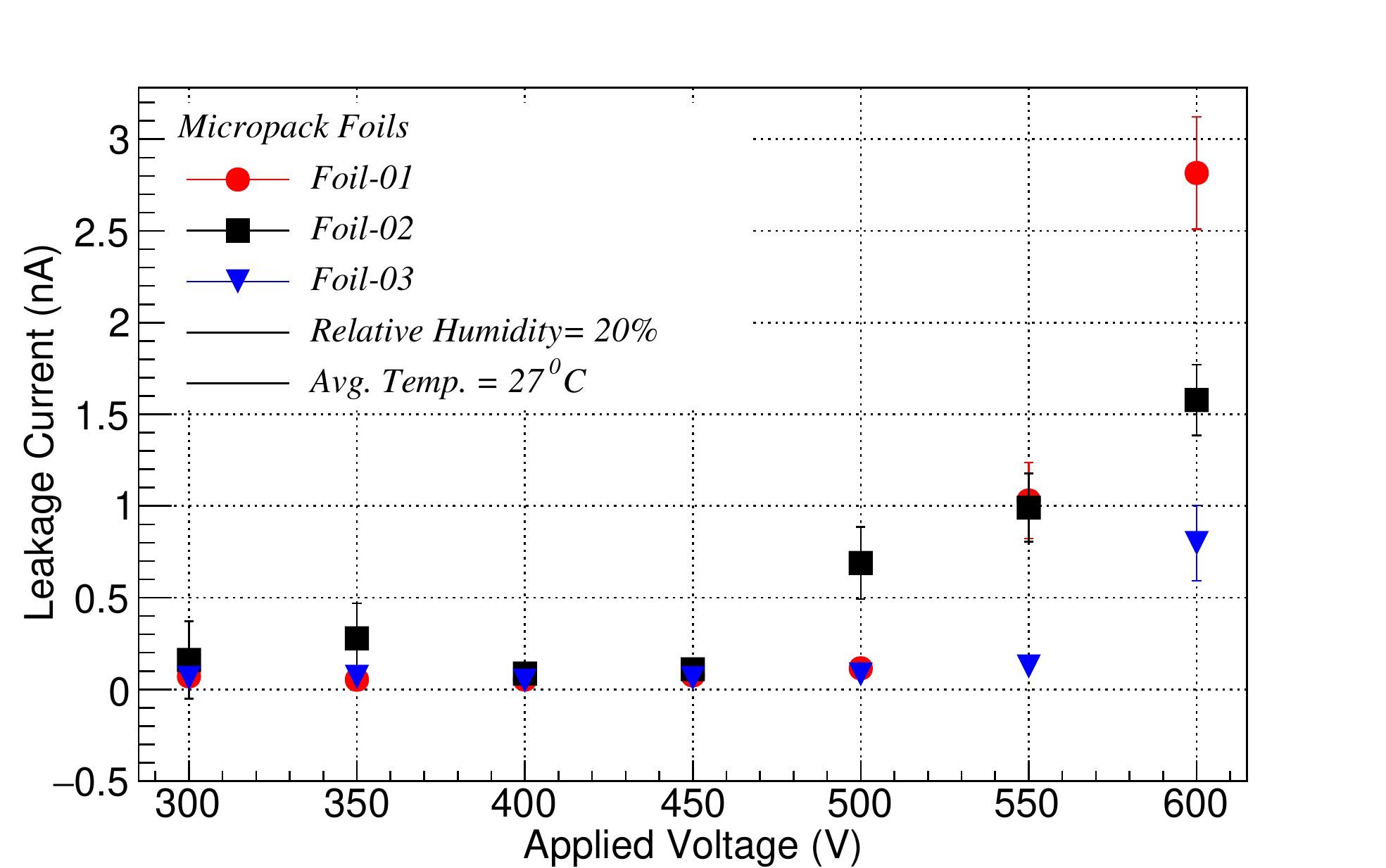}
        \caption{ }
        \label{fig:Indian_foils_H20}
    \end{subfigure}
    \begin{subfigure}[b]{0.46\textwidth}
        \includegraphics[width=7.5cm, height=5.5cm]{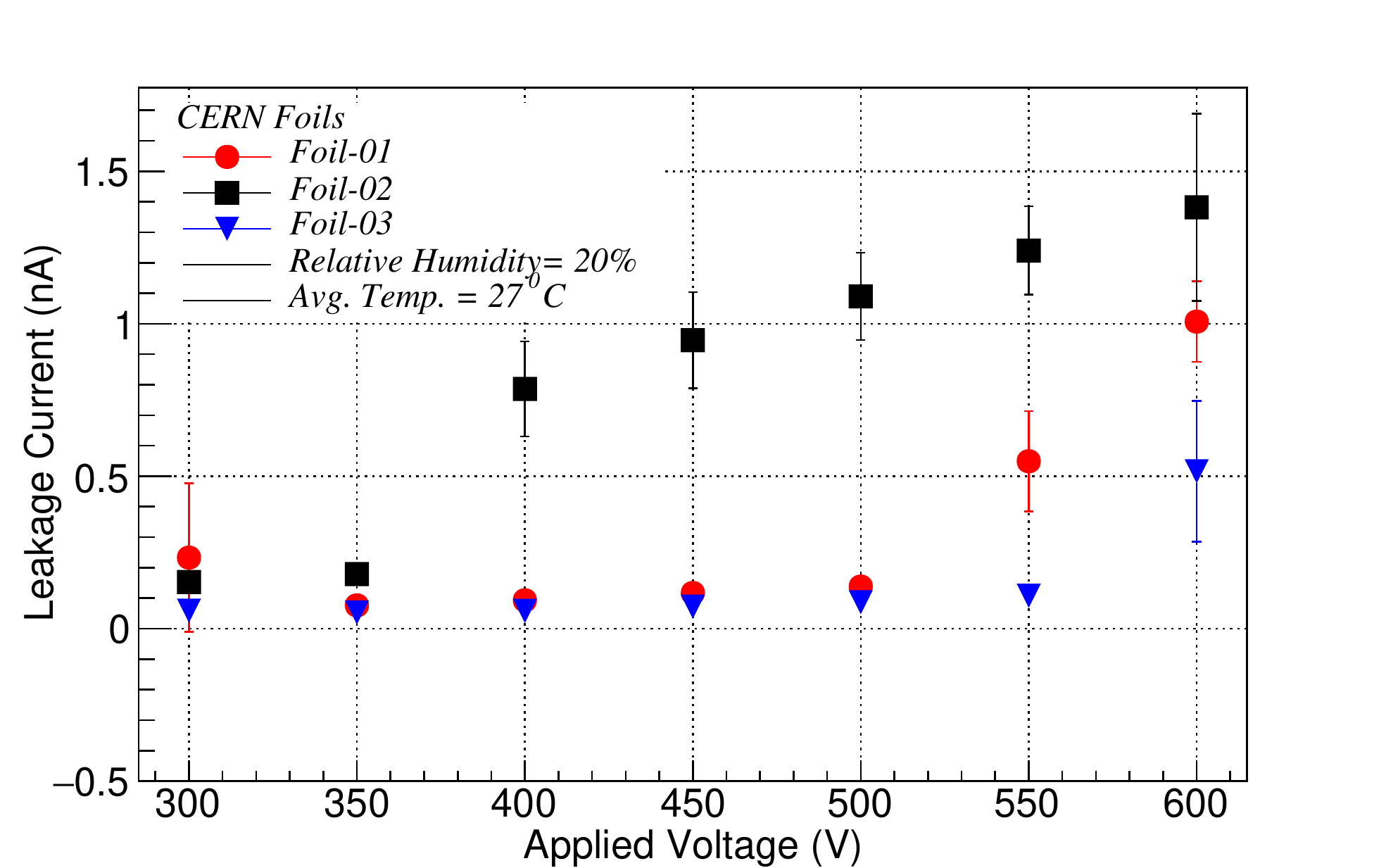} 
        \caption{ }
        \label{fig:CERN_foils}
    \end{subfigure}
   \caption{Leakage Current of (a) Micropack Foils and (b) CERN Foils, at an average temperature of T=27$^{\circ}$C and relative humidity equal to 20\%.} \label{fig:L_01}
\end{figure}

\begin{figure}[!ht]
    \centering
    \begin{subfigure}[b]{0.53\textwidth}
        \includegraphics[width=8cm, height=6cm]{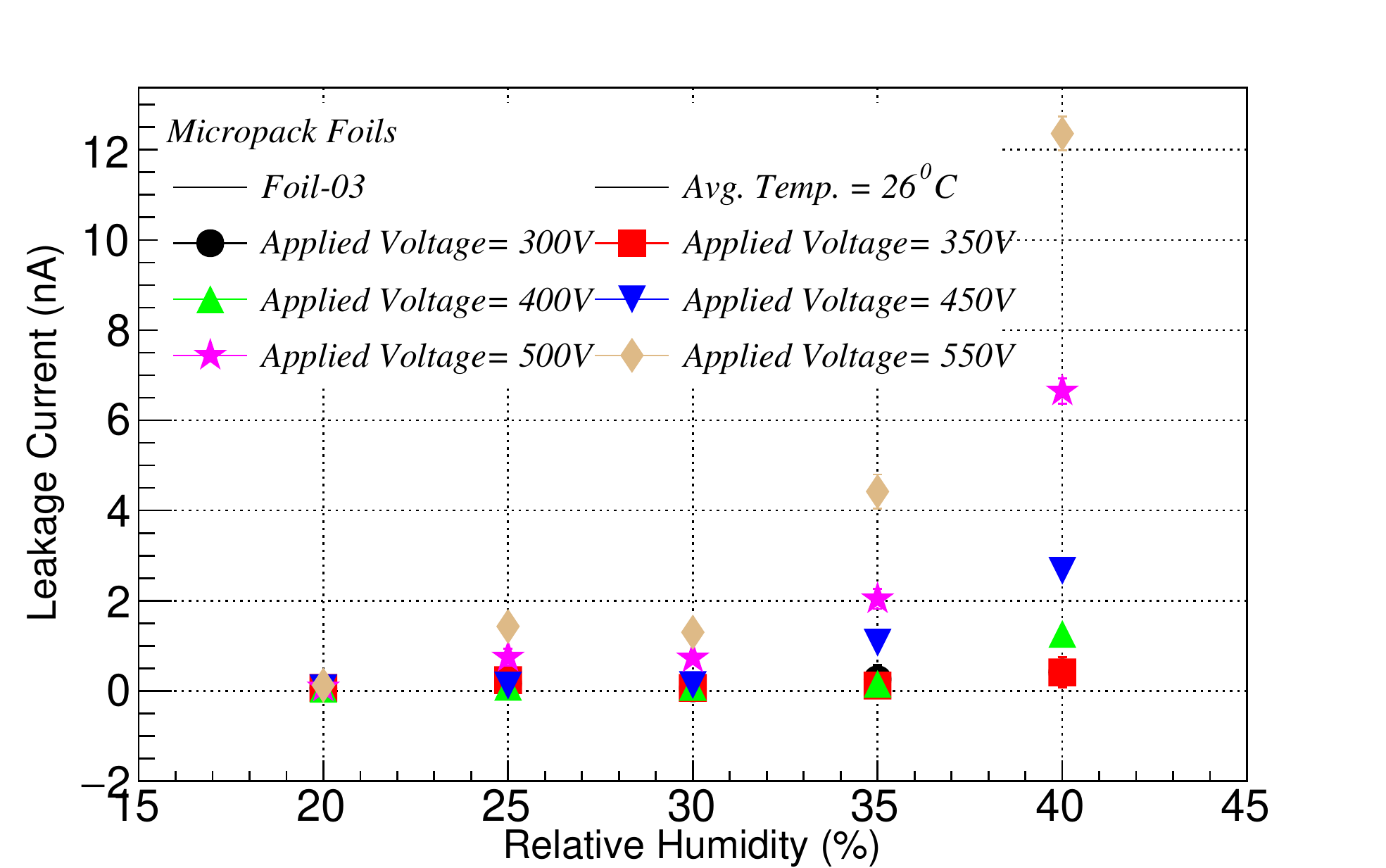}
        \caption{ }
        \label{fig:Indian_foils_IvH}
    \end{subfigure}
    \begin{subfigure}[b]{0.46\textwidth}
        \includegraphics[width=8cm, height=6cm]{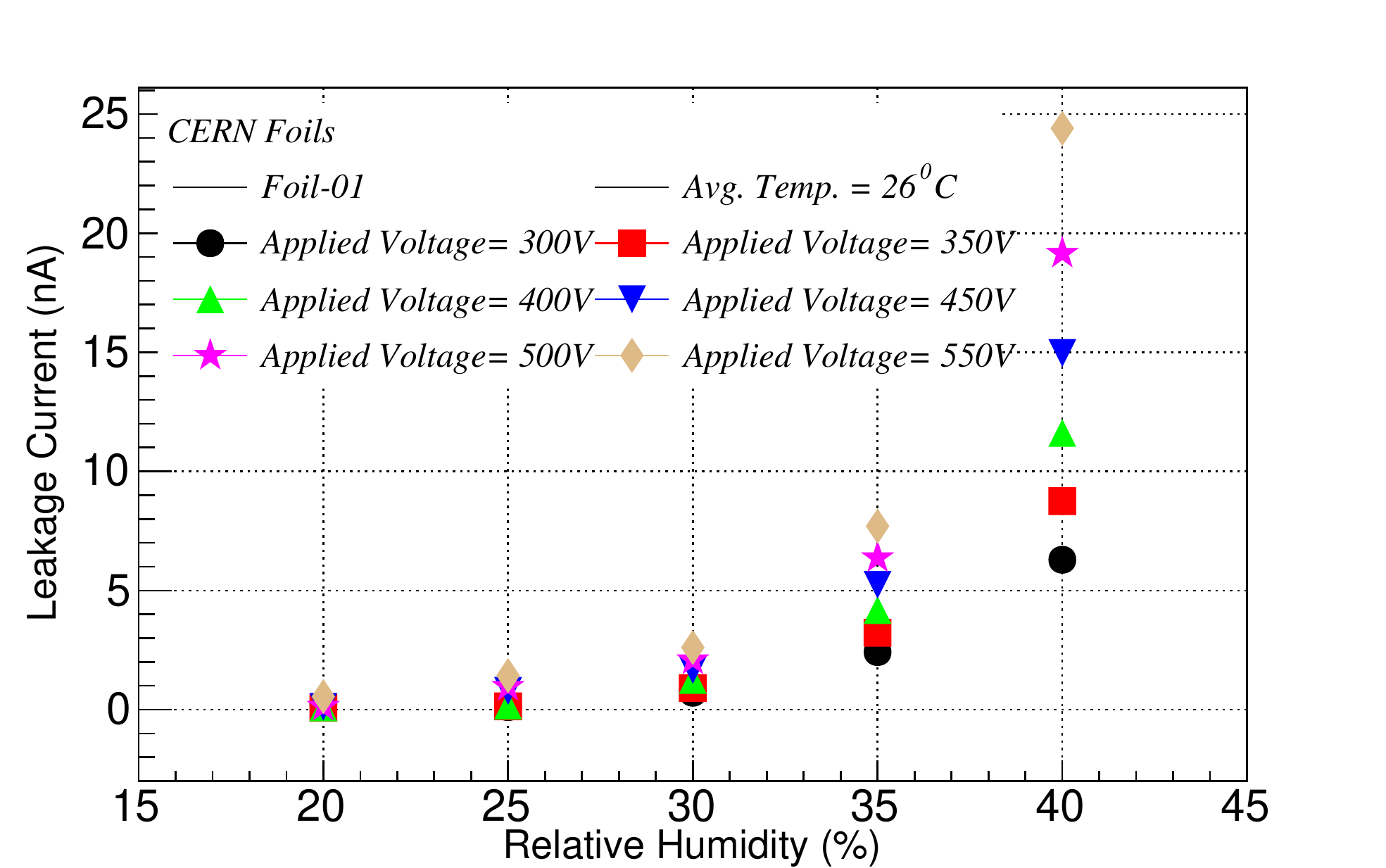}
        \caption{ }
        \label{fig:CERN_foils_IvH}
    \end{subfigure}
   \caption{Leakage Current versus relative humidity taken at different voltages for (a) Micropack Foils (b) CERN foils.} \label{fig:LvH}
\end{figure}
For the better precision in the current measurement, Keithley Electrometer 6517B \cite{thirteen} has been used. The measurement setup consists of a bare GEM foil connected to Keithley 6517B picoammeter interfaced with a computer via a GPIB interface and the Labview program was used to record the measurements as shown in the Figure \ref{fig:Cleaning_Measurement}. The current measurement range was set from 0 to 200 nA, with an accuracy of $\pm$0.2 $\%$. The leakage current thus measured as a function of applied voltage is shown in Figure \ref{fig:L_01} (a) for the Micropack foils. For comparison, the same measurement were also done for foils procured from CERN and the results are shown in the Figure \ref{fig:L_01} (b).  
The Micropack and the CERN foils were found to show similar results under similar ambient conditions. However, as the humidity escalates, the leakage current in CERN foils increases more rapidly compared to the Micropack foils. The maximum current of 12 nA and 25 nA at an applied voltage of 550V, corresponding to the humidity of 40\% has been observed in Micropack and CERN foils respectively. Figure \ref{fig:LvH} shows the leakage current for various applied voltages under different ambient conditions. From the Figure \ref{fig:LvH}, it can be fairly concluded that humidity does have drastic effects on the leakage current measurement. Therefore, the current was also measured in nitrogen environment. Since, Nitrogen is the contamination free standard medium as it is relatively inert and neither reacts with stored materials nor carries moisture. By slowly percolating nitrogen gas into the test enclosure, which in our case was a Plexiglass enclosure in which nitrogen gas was continuously flowing, moisture-laden air was purged out and the current was measured. All the foils showed a current less than 1 nA.

\begin{figure}[!ht]
    \centering
    \begin{subfigure}[b]{0.53\textwidth}
        \includegraphics[width=8cm, height=6cm]{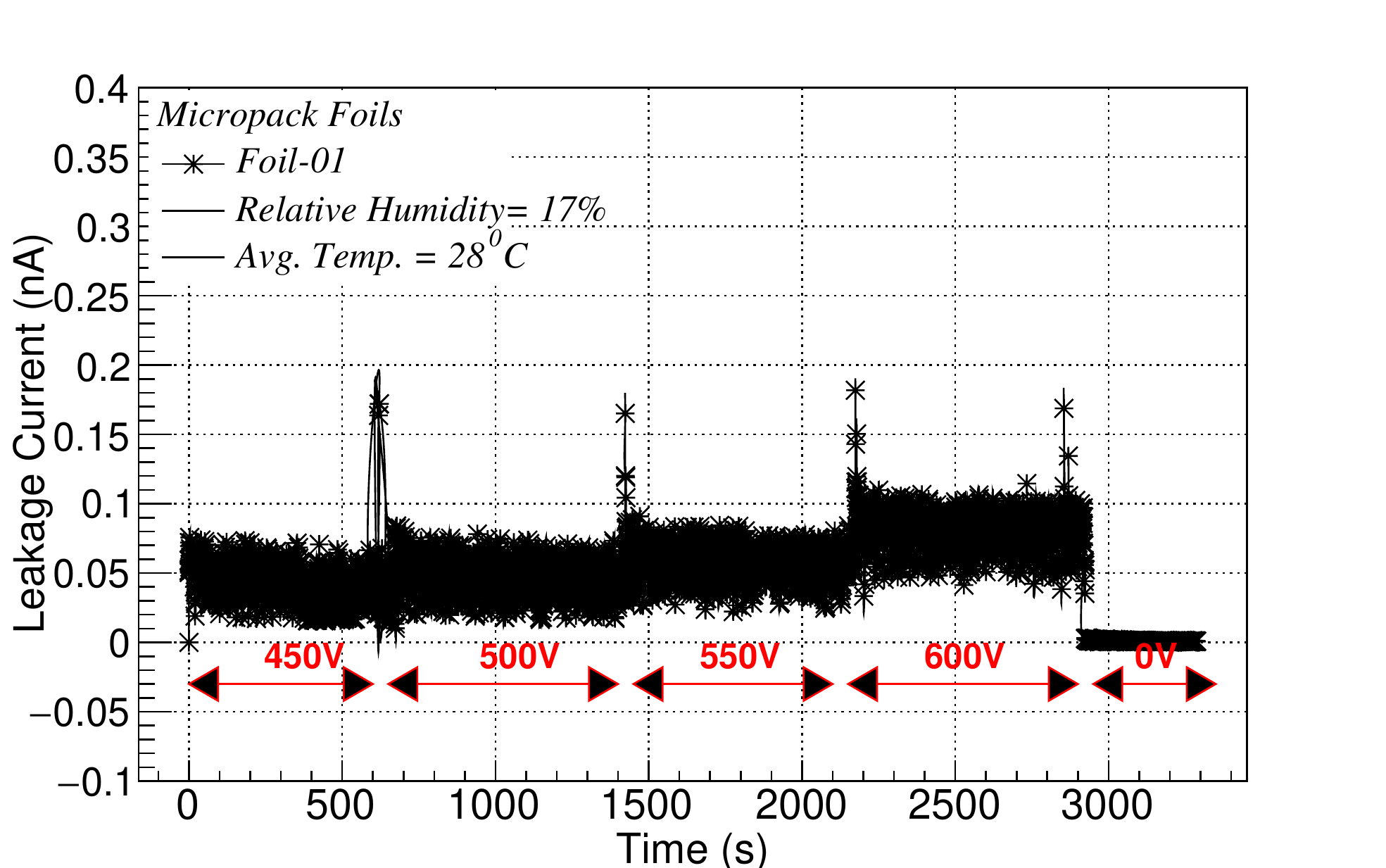}
        \caption{ }
        \label{fig:Micropack_foil_01_lekage_current}
    \end{subfigure}
    \begin{subfigure}[b]{0.46\textwidth}
        \includegraphics[width=8cm, height=6cm]{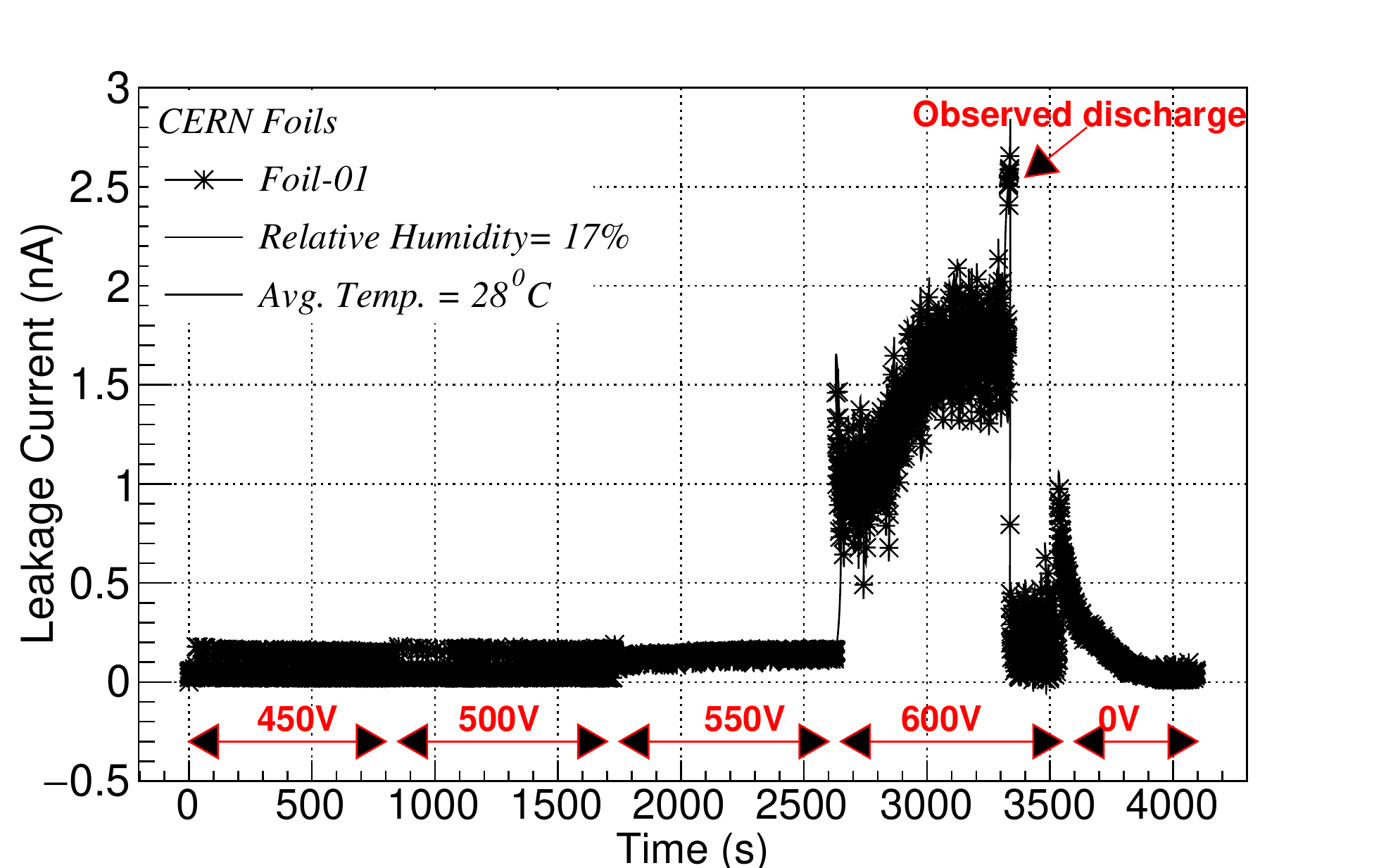}
        \caption{ }
        \label{fig:CERN_foil_01_lekage_current}
    \end{subfigure}
   \caption{QC Long: Leakage Current as a function of time in dry nitrogen environment at different voltage steps with an ambient average temperature of T=28$^{\circ}$C and ambient relative humidity equal to 17\% (a) Micropack foil (b) CERN foil-01.} \label{fig:QC_Long_01}
\end{figure}

\begin{figure}[!ht]
    \centering
    \begin{subfigure}[b]{0.53\textwidth}
        \includegraphics[width=8cm, height=6cm]{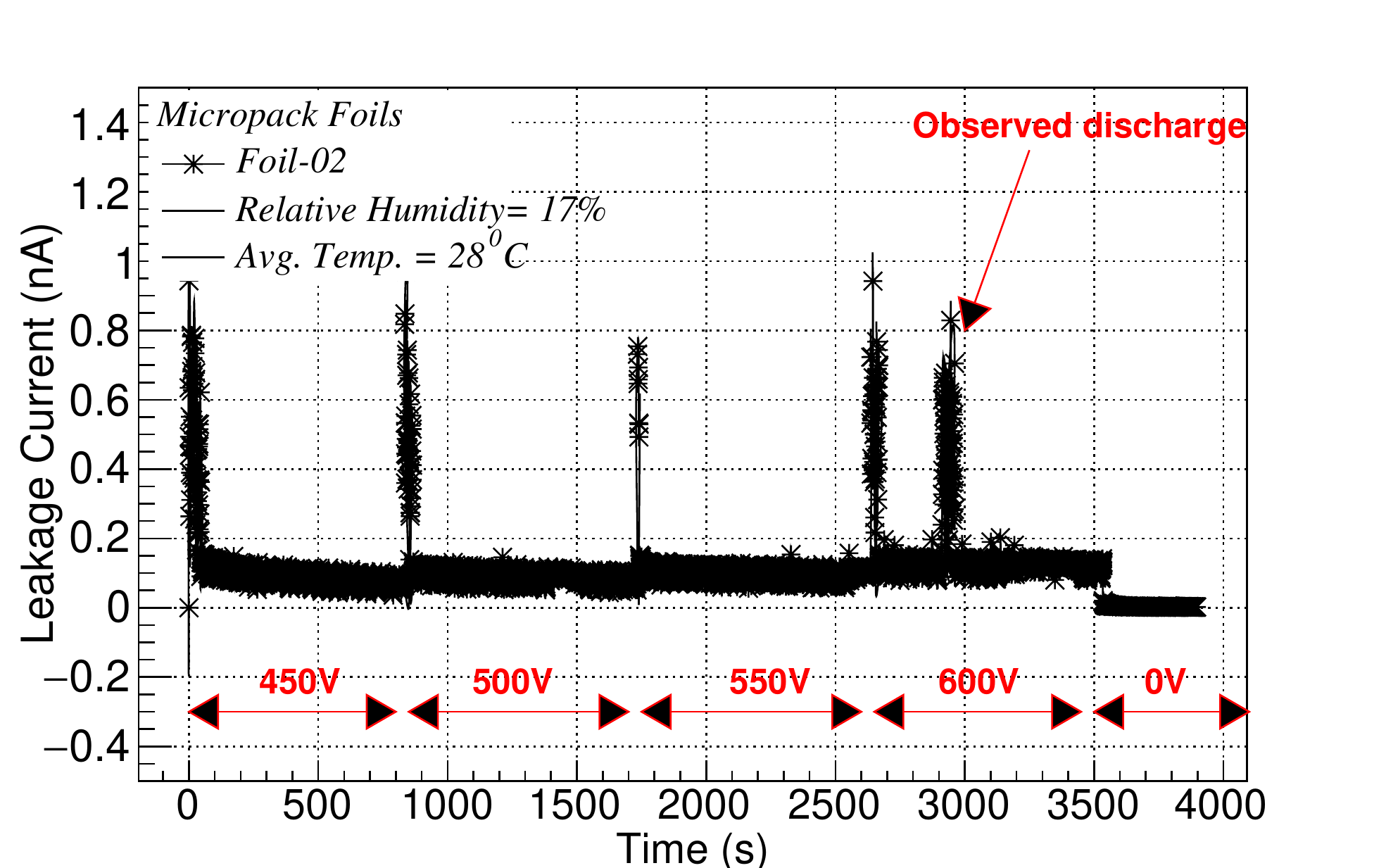}
        \caption{ }
        \label{fig:Micropack_foil_02_lekage_current}
    \end{subfigure}
    \begin{subfigure}[b]{0.46\textwidth}
        \includegraphics[width=8cm, height=6cm]{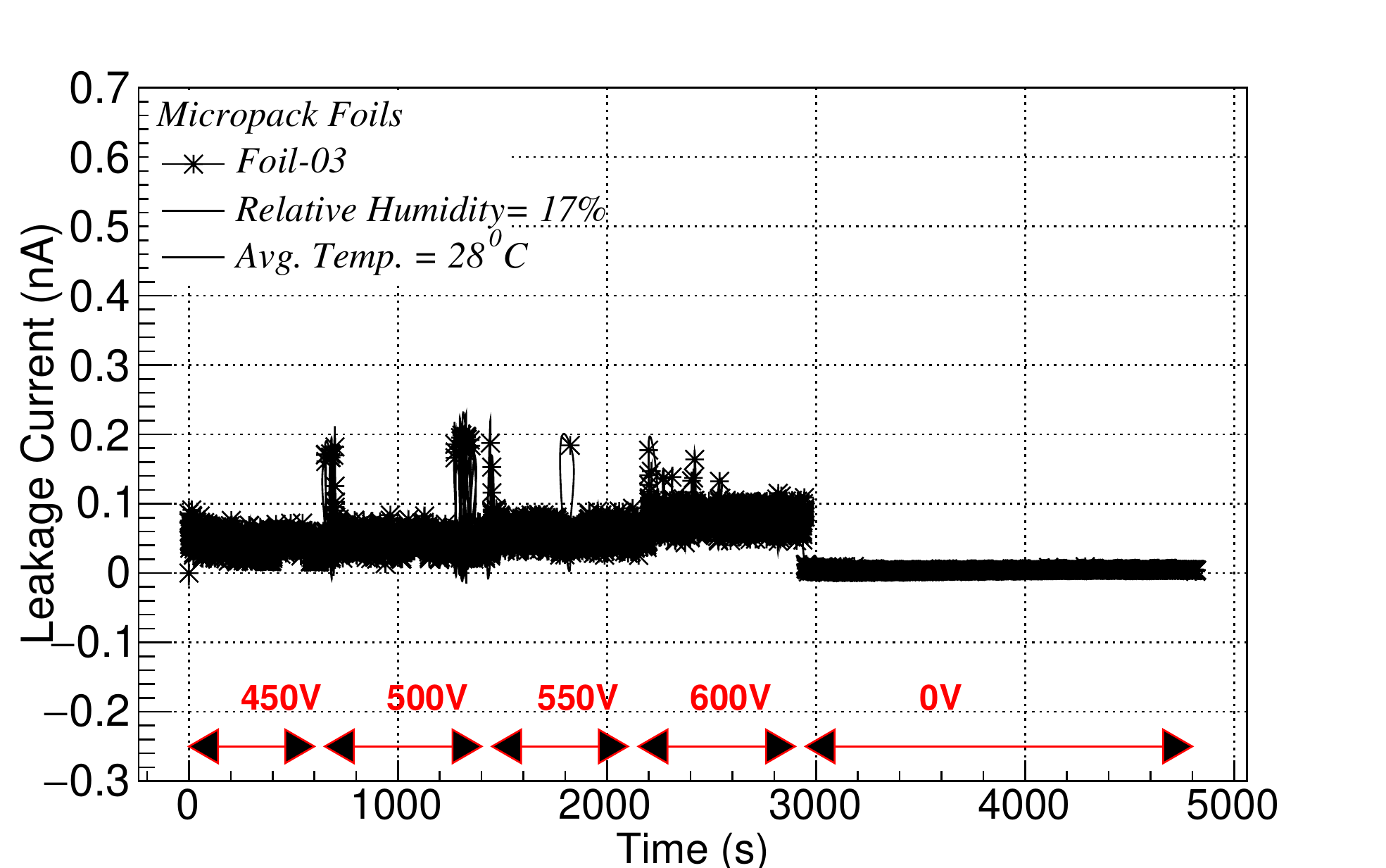}
        \caption{ }
        \label{fig:Micropack_foil_03_lekage_current}
    \end{subfigure}
   \caption{QC Long: Leakage Current as a function of time in dry nitrogen environment at different voltage steps with an ambient average temperature of T=28$^{\circ}$C and relative humidity equal to 17\% (a) Micropack foil-02 (b) Micropack foil-03.} \label{fig:QC_Long_02}
\end{figure}
All the measurements were carried out in the clean room of class 100 installed with a KANOMAX dust particle counter Model 3887 \cite{fourteen} which monitors the particle count. Humidity was controlled by dehumidifier installed in the clean room. The QC long test of the GEM foils were performed by placing foils in a Plexiglass enclosure. After flowing nitrogen continuously for more than two hours, the leakage current was measured in each foil at different voltages in steps of 50V starting from 450V and going up until 600V for time intervals nearly equal to 700s. At 600V, at most two discharges were seen during the time period of around 700s. The corresponding results are shown in Figure \ref{fig:QC_Long_01}. Similar results were obtained for both other foils as shown in Figure \ref{fig:QC_Long_02}.

\section {Conclusion}
GEM foils were produced for the first time in India under the TOT agreement between Micropack Pvt. Ltd. and CERN. Micropack started the preparations for the GEM foil production in India. The first few attempts saw many deviations from the required quality. With further improvements in etching technology and several rounds of iterations, Micropack finally produced a batch of foils which appeared fine from visual inspection and preliminary checks. However, before these foils could be declared fit for applications and technology as reliable, we had to perform the desired quality assessment and characterization for these foils. For this purpose, we performed optical and electrical tests to check the reliability and usability of the foils. Optical tests reveal that the holes are quite uniform with inner and outer diameters of 49.9 $\pm$ 1.6 $\mu$m and 70.01 $\pm$ 2.02 $\mu$m respectively. Here, the quoted errors are the Gaussian one sigma uncertainty on diameter distributions. A current of less than 1 nA has been observed in dry nitrogen environment from electrical measurements and were in agreement with CERN foils. The measured optical and electrical properties of Micropack foils were found to reflect the desired parameters and are at par with the double mask foils produced at CERN. With the successful production of 10 cm $\times$ 10 cm double-mask GEM foils, Micropack has already extended their infrastructure to handle single-mask technology so that larger foils can be produced in order to ease the commercialization of large area GEM foils.
\section*{Acknowledgements}
We would like  to acknowledge the funding agency, Department of Science and Technology (DST), New Delhi for providing generous financial support. We would also like to thank University of Delhi R$\&$D grant to support part of this work. Also, we would like to thank the CERN GEM group, especially Dr. Archana Sharma and Dr. Rui De Oliveira, for providing help during the course of foil development and characterization.

\end{document}